\documentclass[10pt,conference]{IEEEtran}
\usepackage{cite}
\usepackage{amsmath,amssymb,amsfonts}
\usepackage{algorithmic}
\usepackage{graphicx}
\usepackage{textcomp}
\usepackage{xcolor}
\usepackage{graphicx}
\usepackage{subfiles}
\usepackage{array}
\usepackage{amsmath}
\usepackage{algorithm}
\usepackage{hyperref}
\usepackage{multirow}
\usepackage{subfigure}
\usepackage{caption}
\usepackage{flushend}
\usepackage{balance}
\usepackage{diagbox}
\usepackage{ulem}
\usepackage{colortbl}
 \usepackage{enumitem}
 \usepackage{balance}
\usepackage{booktabs}
\usepackage{fancyhdr}
\usepackage[framemethod=TikZ]{mdframed}
\usepackage{tcolorbox}
\makeatletter
\newcommand{\mybox}[1]{%
  \setbox0=\hbox{#1}%
  \setlength{\@tempdima}{\dimexpr\wd0+13pt}%
  \begin{tcolorbox}[boxrule=0.5pt, colback=gray!10, arc=4pt,
      left=6pt,right=6pt,top=6pt,bottom=6pt,boxsep=0pt]
    #1
  \end{tcolorbox}
}

\usepackage{makecell}

\usepackage{float}

\newcommand{\tool}{\textsc{DOME}}

\newcommand{\ltool}{\textsc{COIN}}

\def\BibTeX{{\rm B\kern-.05em{\sc i\kern-.025em b}\kern-.08em
    T\kern-.1667em\lower.7ex\hbox{E}\kern-.125emX}}
\begin{document}

\title{Developer-Intent Driven Code Comment Generation\\
}

\author{%
  \IEEEauthorblockN{%
    Fangwen Mu\IEEEauthorrefmark{1}\IEEEauthorrefmark{2}\IEEEauthorrefmark{3},
    Xiao Chen\IEEEauthorrefmark{1}\IEEEauthorrefmark{2}\IEEEauthorrefmark{3},
    Lin Shi\IEEEauthorrefmark{1}\IEEEauthorrefmark{2}\IEEEauthorrefmark{3}\textsuperscript{\textsection},
    Song Wang\IEEEauthorrefmark{5},
    Qing Wang\IEEEauthorrefmark{1}\IEEEauthorrefmark{2}\IEEEauthorrefmark{3}%
  }%
  \IEEEauthorblockA{\IEEEauthorrefmark{1} State Key Laboratory of Intelligent Game, Beijing, China}%
  \IEEEauthorblockA{\IEEEauthorrefmark{2} Science and Technology on Integrated Information System Laboratory, \\ Institute of Software Chinese Academy of Sciences, Beijing, China}%
  \IEEEauthorblockA{\IEEEauthorrefmark{3} University of Chinese Academy of Sciences, Beijing, China}%
  \IEEEauthorblockA{\IEEEauthorrefmark{5} Lassonde School of Engineering, York University, Toronto, Canada}%
  \IEEEauthorblockA{ \{fangwen2020, chenxiao2021, shilin, wq\}@iscas.ac.cn, wangsong@yorku.ca}
}

\maketitle
\begingroup\renewcommand\thefootnote{\textsection}
\footnotetext{Corresponding author.}
\endgroup

\begin{abstract}
Existing automatic code comment generators mainly focus on producing a general description of functionality for a given code snippet without considering developer intentions. However, in real-world practice, comments are complicated, which often contain information reflecting various intentions of developers, e.g., functionality summarization, design rationale, implementation details, code properties, etc. 
To bridge the gap between automatic code comment generation and real-world comment practice, we define Developer-Intent Driven Code Comment Generation, which can generate {intent-aware} comments for the same source code with different intents. 
To tackle this challenging task, we propose DOME, an approach that utilizes Intent-guided Selective Attention to explicitly select intent-relevant information from the source code, and produces various comments reflecting different intents. 
Our approach is evaluated on two real-world Java datasets, and the experimental results show that our approach outperforms the state-of-the-art baselines. 
A human evaluation also confirms the significant potential of applying {\tool} in practical usage, enabling developers to comment code effectively according to their own needs.
\end{abstract}


\begin{IEEEkeywords}
Code Comment Generation, Intent-Controllable Comment Generation, Automated Comment-Intent Labeling
\end{IEEEkeywords}

\section{Introduction}
\label{sec:introduction}



\begin{figure}[t]
\centering
\includegraphics[width=0.9\columnwidth]{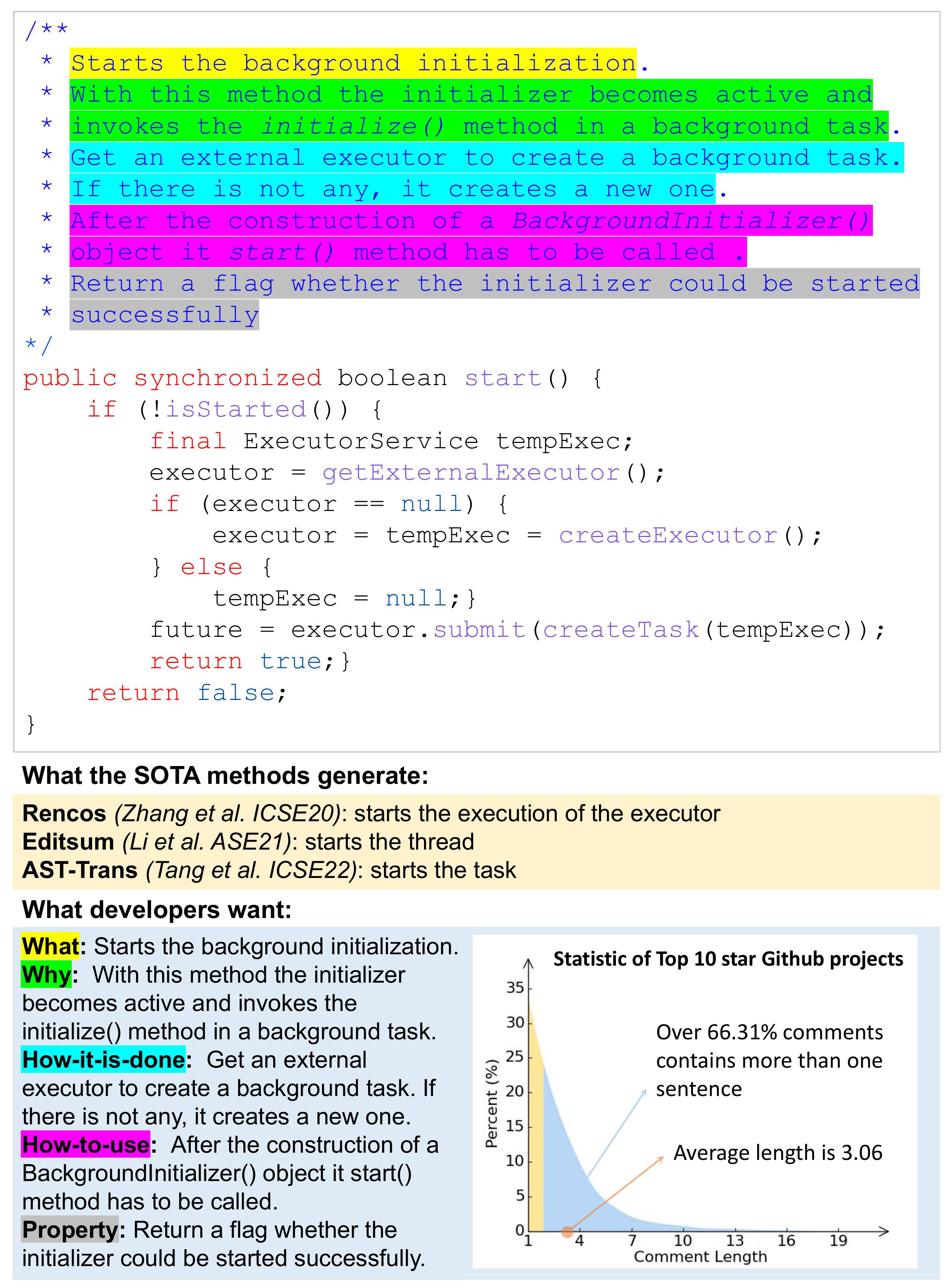}
\caption{A motivation example of intent driven code comment generation.}
\vspace{-0.7cm}
\label{fig:motivation}
\end{figure}

Code comment generation concerns the production of a concise and fluent description of source code that facilitates software development and maintenance by enabling developers to comprehend, ideate, and document code effectively.
Typically comment generation methods model the input code and output comment as a one-to-one mapping without considering developers' intents. Whereas, a code snippet is often associated with multiple comments reflecting different intents, which is a one-to-many mapping.
As the example shown in Figure~\ref{fig:motivation}, the human-writing comment of the method \textit{start()} consists of five sentences that reflect the different intent of the developer. The first sentence summarizes the overall functionality of the code, the second sentence explains the design rationale, and the 3rd-5th sentences describe the implementation details, the usage, and the property of the code, respectively. 
However, the comments automatically generated by the three state-of-the-art (SOTA) methods only describe the functionality of the method \textit{start()}. 
Furthermore, we analyzed the methods of the top 10 Java projects with the most stars from GitHub, and found that over 66.31\% comments contain more than one sentence. 
At the same time, we found that each comment involves 2.81 different intents on average by manually reviewing 100 comments. 
The observation indicates that the one-to-one code comment generation can hardly fulfill practical needs.

Thus it is appealing and important to develop an approach to generate comments which can satisfy various intents. To bridge the gap, we define developer-intent driven code comment generation, which aims to produce comments that are coherent with the given intents, i.e., \emph{what}, \emph{why}, \emph{how-to-use}, \emph{how-it-is-done}, and \emph{property}, following previous work~\cite{DBLP:journals/tosem/ChenXHLL21}\cite{DBLP:conf/icse/ZhaiXSTPMXZTZ20}.

In practice, developers may focus on a particular aspect instead of a full description of the code when writing different kinds of comments.
For example, when the developers aim to describe the `How-it-is-done' of the \textit{start()} method as shown in \ref{fig:motivation}, they would pay more attention to the middle part of the code, which contains more running logic and implementation details. When the developers aim to describe the `Property', they would pay more attention to the return value at the end and the parameter type of the method at the beginning. 

In light of this, we propose {\tool}, a Developer-intent driven cOde coMment gEneration approach that can generate various comments for one code snippet under different intents by leveraging intent-guided selective attention. Specifically, {\tool} consists of three main components: an Exemplar Retriever, an Encoder Layer, and a Decoder Layer. Given a code snippet and an intent category, the Exemplar Retriever first selects the code-comment pairs with the same intent as the given intent from the pre-defined corpus as the retrieval corpus. Then, it employs the DPR model~\cite{DBLP:conf/emnlp/KarpukhinOMLWEC20} to retrieve the most similar comment from the retrieval corpus and treat it as the exemplar. 
Next, we input the code snippet, intent category, and the exemplar into the Encoder Layer to encode them into semantic representations. Finally, the decoder layer equipped with intent-guided selective attention is guided by the given intent to select the most relevant information from the semantic representations to generate an intent-aware comment. 

Furthermore, since training and evaluating {\tool} require a large volume of labeled comment-intent data, we develop a COmment-INtent labeling tool, named COIN, to support the automatic annotation of comment intents for the code-comment dataset. Specifically, we first sample a total of 20K code-comment data from two large-scale Java datasets, and manually annotate these data to train {\ltool}, which achieves the high performance of 89.6\% Macro-F1 on average.
The well-trained {\ltool} is then utilized to automatically annotate the large-scale code-comment corpus that will be used to train our comment generation model {\tool}.

To evaluate our approach, we conduct experiments on two real-world datasets in Funcom \cite{leclair2019neural} and TLC \cite{DBLP:conf/ijcai/HuLXLLJ18}, and the results show that our approach outperforms the state-of-the-art (SOTA) baselines by 25.66\%, 16.59\%, and 18.38\% with respect to BLEU-4, ROUGE-L, and METEOR on Funcom dataset. On TLC dataset, {\tool} improves the performance on BLEU-4, ROUGE-L, and METEOR by 10.06\%, 11.09\%, and 14.93\%, respectively. We also conduct a human evaluation to assess the generated comments on three aspects: accuracy, adequacy, and naturalness, showing that {\tool} can generate useful and relevant comments.

Our main contributions are outlined as follows:
\begin{itemize}
    \item \textbf{Technique}: a novel comment generation model, named {\tool}, which utilizes the intent-guided selective attention to explicitly select relevant information from source code based on the given intent for generating comments. To the best of our knowledge, this is the first work that incorporates developer intents in comment generation. 
    \item \textbf{Labeling Tool}: an automated comment-intent labeling tool, named {\ltool}, which helps  build high-quality intent-annotated code comment datasets.
    \item \textbf{Evaluation}: an experimental evaluation of {\tool} against state-of-the-art baselines, which shows that {\tool} outperforms all baselines, together with a human evaluation, which further confirms the significant potential of applying {\tool} in real-world practice, for enabling developers to comment code effectively according to their own needs. 
    \item \textbf{Data}: publicly accessible dataset and source code~\cite{website} to facilitate the replication of our study and its application in extensive contexts.
\end{itemize}

\section{Background and Problem Definition}
\label{sec:background}

\subsection{Taxonomy of Comment Intent}

In this work, we use the intent taxonomy of code comments proposed by~\cite{DBLP:journals/tosem/ChenXHLL21}, which consists of six categories, i.e.,  \emph{what, why, how-to-use, how-it-is-done, property}, and \emph{others}, as described in Table \ref{table:label_description}. {Note that, since the \emph{others} comments are defined as the unspecific and ambiguous comments}, we consider the code-comment pairs with the intent of \emph{others} as noisy data, and remove them if identified.

\begin{table}[hbtp!]
\centering
\caption{The intent taxonomy of code comments\cite{DBLP:journals/tosem/ChenXHLL21}}
\label{table:label_description}
\begin{tabular}{>{\centering}m{0.1\columnwidth}>{\raggedright}m{0.4\columnwidth}m{0.3\columnwidth}}

\hline
Category       & Description    & Example    \\
\hline
What           & Describes the functionality of a method                                        & \textit{``A helper function that process the stack.''}                                                \\
Why            & Explains the reason why a method is provided or the design rationale of the method      & \textit{``Get a copy of the map (for diagnostics)''}                                                  \\
How-to-use     & Describes the usage or the expected set-up of using a method                            & \textit{``Should be called before the object is used''}                                               \\
How-it-is-done & Describes the implementation details of a method                                       & \textit{``Convert the byte{[}{]} to a secret key''}                                                   \\
Property       & Asserts properties of a method including pre-conditions  or post-conditions of a method &\textit{ ``Wait until seqno is greater than or equal to the desired value or we exceed the timeout.''} \\
Others         & Unspecified or ambiguous comments                                                         & \textit{``The implementation is awesome.''}  \\                            
\hline
\end{tabular}
\end{table}

\subsection{Transformer}
In this work, we use the Transformer \cite{DBLP:conf/nips/VaswaniSPUJGKP17} as the backbone to construct  {\tool}. Transformer is a relatively popular model in recent years. It has achieved promising results in many fields, such as machine translation \cite{DBLP:conf/acl/WangLXZLWC19} and text summarization \cite{DBLP:conf/emnlp/LiuL19}. Transformer follows the encoder-decoder framework with stacked encoder blocks and decoder blocks. There are two main layers in each encoder block and decoder block, i.e., a Multi-head Attention Layer (MHA) and a Feed-Forward Network (FFN). The residual connection is employed around each layer, followed by layer normalization (Norm) \cite{DBLP:journals/corr/BaKH16}. Since Transformer removes the recurrence mechanism, it cannot utilize the order information of input tokens directly. Therefore, positional encoding (PE) is used in Transformer to provide the position information of each token.

\begin{figure*}[t]
\centering
\includegraphics[width=0.95\textwidth]{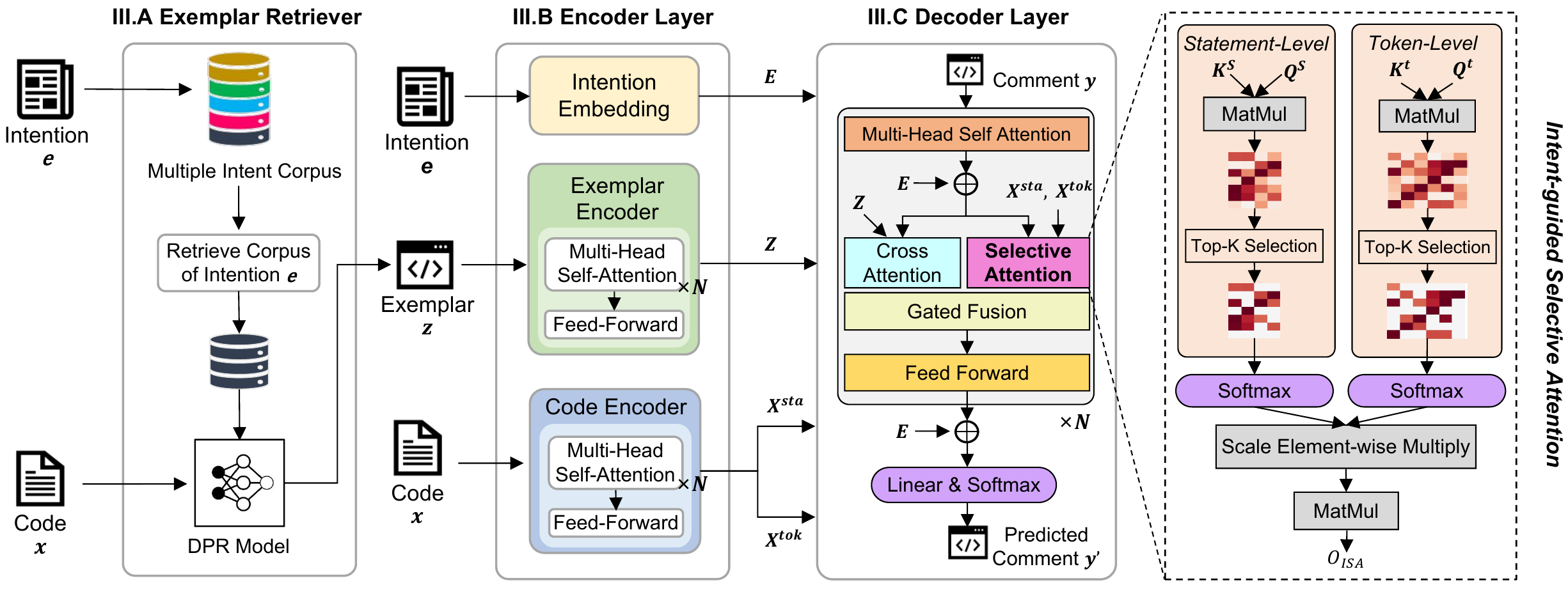}
\caption{The architecture of {\tool}}
\label{fig:approach}
\end{figure*}

\subsection{Problem Definition}
The developer-intent driven code comment generation task is formulated as follows: Given a code snippet $x$ and an intent category $e$ of the comment to be generated, the goal of the task is to generate a comment $y$ that reflects the intent $e$. Essentially, the model learns to estimates the probability: $P(y|x, e) = \prod_{i=1} P(y_i|y_{<i}, x, e)$ when training, and generate the prediction comment $y'$ that maximizes the conditional likelihood $y'={argmax}_{y'}P(y'|x, e)$ when inference. 

\section{Approach}
\label{sec:method}

Figure \ref{fig:approach} illustrates the overview of {\tool}, which consists of three main components:
(1) \textbf{Exemplar Retriever}, for retrieving the most similar comment as the exemplar, which can provide essential clues about linguistic patterns and expressions;
(2) \textbf{Encoder Layer}, for encoding the source code, retrieved exemplar, and target intent into semantic representations;
and (3) \textbf{Decoder Layer}, for leveraging the intent-guided selective attention to extract the most relevant information from the semantic representations and generating the intent-aware comments. 

\subsection{Exemplar Retriever}
{Suppose we have a multiple intent corpus $D$ that consists of triples 
$<code_i, comment_i, intent_i>$, where $comment_i$ is the comment for code snippet $code_i$ under the intent category $intent_i$.}
Given a code snippet $x$ and an intent category $e$, 
we first collect triplets with the same intent as the given category $e$ from the corpus $D$, and take them as the retrieval corpus $D_e$. This step is simple yet effective, as 
(1) intuitively, comments with different intents are different in content and expression, and treating them as exemplars may mislead the model to generate comments that are irrelevant to the target intent $e$. 
(2) it can largely reduce the number of candidate triples in the retrieval corpus, especially for large-scale datasets, 
so as to improve the speed of training and inference. 

Then, we use the retrieval techniques to match the most similar comment from the retrieval corpus based on the given code $x$.
{In previous work~\cite{DBLP:conf/icse/ZhangW00020,DBLP:conf/kbse/LiL000J21,DBLP:conf/kbse/WeiLLXJ20}, the traditional term-based retrieval techniques (e.g., TF-IDF\cite{DBLP:journals/jd/Jones04} and BM25\cite{DBLP:journals/ftir/RobertsonZ09}) have been widely used.} 
Although the term-based retrieval methods have the advantages of time-saving and convenience, it has been pointed out that they may cause the model to fail to converge \cite{DBLP:conf/kbse/LiL000J21} or hurt the model performance \cite{DBLP:conf/iclr/LiuCXS021} since they cannot exploit semantic-level features of the code and comments, and are prone to retrieval of dissimilar data. To alleviate this problem, we employ the Dense Passage Retriever (DPR) \cite{DBLP:conf/emnlp/KarpukhinOMLWEC20} model as the retriever. 
DPR is the SOTA technique for open-domain question answering, which contains two encoders that encode queries and passages into dense vector representations, respectively. It can leverage the semantic-level information of queries and passages, and measure their similarity score by calculating the dot product. 
The DPR model has been shown to be effective in code search and code comment generation tasks \cite{DBLP:conf/emnlp/ParvezACRC21}. We adopt
the pretrained DPR model provided by \cite{DBLP:conf/emnlp/ParvezACRC21} to retrieve the example $z$. 


\subsection{The Encoder Layer}
Once we have an exemplar $z$, we feed it into the Encoder Layer together with the code $x$ and the intent $e$.
As shown in figure \ref{fig:approach}, the Encoder Layer consists of one intent embedding layer and two different encoders (i.e., code encoder and exemplar encoder). 
The intent embedding layer is utilized to capture the high-level abstraction of intent expressions. The code encoder and exemplar encoder aim to extract the semantic features from the code snippet and the retrieved exemplar, respectively.
We construct the two encoders by following the structure of the vanilla Transformer Encoder \cite{DBLP:conf/nips/VaswaniSPUJGKP17} that we have introduced in Section~\ref{sec:background}.
The only difference is that our code encoder outputs two-level representation sequences for tokens and statements, respectively. There are two reasons that we take the additional statement-level information into consideration: First, statements are essential units for carrying source code semantics \cite{DBLP:conf/icse/ZhangWZ0WL19}. Second, it is used to calculate the intent-guided selective attention for extracting intent-relevant semantic features, which will be described in Section \ref{selective attention}.

\subsubsection{Code Encoder}
Assume a code snippet $x=[x_1, x_2, ..., x_L]$ contains $L$ statements, and the $l$-th statement is denoted as $[x_{l,1}, x_{l,2},...,x_{l,M}]$, where $x_{l,i}$ is the $i^{th}$ token.
{When preparing the embedding sequences, we combine all the statements of the given code snippet into one sequence, rather than input each statement into the model separately. It is mainly because that statements in different sequences can hardly share and convey information to each other.}
{To make the code encoder understand the end of one statement and the start of another statement in the same sequence,}
we insert a special token [SEP] after the end of each statement, so the augmented statement $x_l = [x_{l,1}, x_{l,2},...,x_{l,M+1}]$, where $x_{l,M+1}$ is the [SEP]. 
The code snippet $x=[x_{1,1}, ..., x_{l,m},...,x_{L,M+1}]$ is first converted into a sequence of $d$ dimensional embeddings $\overrightarrow{x}\in\mathbb{R}^{\scriptscriptstyle(M+1)L\times d}$ via a token and position embedding layer.
Then, we input the embedding sequence $\overrightarrow{x}$ into the $N$ identical encoder blocks to calculate the token-level representations. For $n$-th block of the code encoder, suppose that the input is $H^{n-1}$, the output $H^n$ is calculated as follows:
\begin{equation}
H^{n}_1 = {\rm Norm}\bigg(H^{n-1} + {\rm MHA}(H^{n-1},H^{n-1},H^{n-1})\bigg) \label{eq:first_layer}
\end{equation}
\begin{equation}
H^n = {\rm Norm}\bigg(H^{n}_1 + {\rm FFN}(H^{n}_1)\bigg) \label{eq:last_layer}
\end{equation}
where $H^{n}_1$ is the hidden states of the first layer in $n$-th encoder block. Initially, the embedding sequence $\overrightarrow{x}$ is fed into the first block, and the $N$-{th} block outputs the final token representation $X^{tok}\in\mathbb{R}^{\scriptscriptstyle(M+1)L\times d}$.
Next, we perform the MaxPooling on the token representation in each statement to compute the representation of that statement:
\begin{equation}
X^{sta}_l = {\rm MaxPooling}([X^{tok}_{l,1}, X^{tok}_{l,2}, ..., X^{tok}_{l,M+1}]) \label{eq:mean_pooling}
\end{equation}

We concatenate each statement representation to obtain the statement-level representation sequence $X^{sta}\in\mathbb{R}^{\scriptscriptstyle L\times d}$.

\subsubsection{Exemplar Encoder}
{We construct the exemplar encoder by using the same structure as the code encoder but with different parameters.}
Similar to the code encoder, the exemplar encoder embeds the retrieved exemplar $z=[z_1, z_2, ..., z_T]$ into the sequence of embeddings $\overrightarrow{z}\in\mathbb{R}^{\scriptscriptstyle T\times d}$. Then, the exemplar representation $Z\in\mathbb{R}^{\scriptscriptstyle T\times d}$ can be computed via the equation (\ref{eq:first_layer}) and (\ref{eq:last_layer}).
 
\subsubsection{Intent Embedding}
Since the comment intents provide a high-level semantic abstraction of the comment, we take the intents as additional input and map them into the dense semantic vectors. 
For each intent $e$, we use an embedding matrix to map it into the intent embedding vector $E$, and then update the parameters of the embedding matrix through training. $E$ will be utilized to guide the decoder to select intent-relevant information from the outputs of the encoders.

\subsection{The Decoder Layer}
The decoder layer aims to produce the intent-aware comment by explicitly capturing the important clues from the encoder outputs based on the intent embedding.
As shown in Figure \ref{fig:approach}, the decoder is composed of a stack of $N$ identical decoder blocks, and each block consists of three layers where the first and the last layers are the same as those in the encoder. The additional layer contains an Intent-guided Selective Attention (ISA) and a Multi-Head Cross Attention followed by a Gated Fusion layer.
In this section, we first introduce the decoding process of the decoder and then describe the details of the Intent-guided Selective Attention.

\subsubsection{Decoding Process}
Given the representation sequences $X^{tok}$, $X^{sta}$, $Z$ and the intent embedding $E$, the $n$-{th} decoder block first gets the output of the first layer $S^n_1$ via Eq. (\ref{eq:first_layer}). 
Then, in the second layer, the block concatenates the hidden states $S^n_1$ and intent embedding $E$ as the query vector:
\begin{equation}
    Q^n_1 = [S^n_1~;~E] \label{eq:concat}
\end{equation}
where $[;]$ denotes concatenation operation.
Next, it captures information from the source code and exemplar by performing ISA over the token-level and statement-level representations and MHA over the exemplar representation, respectively:
\begin{equation}
    O^n_{\rm ISA} = {\rm ISA} (Q^n_1, X^{tok}, X^{sta})
\end{equation}
\begin{equation}
    O^n_{\rm MHA} = {\rm MHA} (Q^n_1, Z, Z)
\end{equation}
With the equation (\ref{eq:concat}), the model can obtain the intent semantics and focus more on the information that related to the intent.
To effectively leverage the information from the source side, we utilize the gate mechanism~\cite{DBLP:journals/neco/HochreiterS97} to adaptively incorporate the $O^n_{\rm ISA}$ containing source code features and the $O^n_{\rm MHA}$ containing exemplar features:
\begin{align}
&\beta = {\rm Sigmoid}(W^\mathrm{\emph{T}}_{gate}[O^n_{\rm ISA}~;~O^n_{\rm MHA}]) \\   
&S^n_{2} = \beta\cdot O^n_{\rm ISA}+ (1-\beta)\cdot O^n_{\rm MHA} 
\end{align}
where $\beta$ is the degree of integration between source code and exemplar. A larger value of the $\beta$ (ranges from 0 to 1) may indicate that the retrieved exemplar is semantically different from the source code, and the model should pay more attention to the source code. $W_{gate}$ is a trainable parameter matrix, and $S^n_{2}$ is the hidden states of the second layer.
Then, according to Eq. (\ref{eq:last_layer}), the $n$-{th} block uses the $S^n_{2}$ to compute the output of the last layer $S^n$. 
After the calculation of $N$ decoder blocks, the decoder gets the hidden states of the last decoder block $S$. 

For the $i$-th decoding step, the probability of $i$-{th} token $y'_i$ can be calculated by projecting the concatenation of the state $s_i$ and intent Embedding $E$ via a linear layer followed by a Softmax function.
\begin{equation}
p(y'_i|y'_1,~y'_2,~...,~y'_{i-1}) = {\rm Softmax}(W_o^\mathrm{\emph{T}}[s_j;E] + b_o) 
\end{equation}
where $W_o$ is the parameter matrix and $b_o$ is the bias. Ultimately, we use the Argmax function to generate the prediction comment $y'$.
\begin{equation}
y' = {\rm Argmax}([p(y'_1)~;~\cdots~;~p(y'_i)~;~\cdots])
\end{equation}

\subsubsection{Intent-guided Selective Attention (ISA)}
\label{selective attention}

{To make the decoder focus on a particular aspect instead of a complete description of the code when generating different kinds of comments, we propose intent-guided selective attention that enables the model to catch the intent-relevant information and ignore the irrelevant noise.}
Our proposed attention variant contains three steps: (1) statement-level attention selection, (2) token-level attention selection, and (3) combining attentions.

\textbf{Statement-level Attention Selection.}
In this step, our goal is to select the most relevant statements based on the given intent. The inputs are the intent embedding $E$, statement representation $X^{sta}$, and query vector $Q^n_1$ which is the concatenation of the hidden states $S^n_1$ and intent embedding $E$.
We treat the $Q^n_1$ as the query, $X^{sta}$ as the key and the value, and perform a linear projection on them: 
\begin{equation}
\mathcal{Q}_s = Q^n_1W^{\mathcal{Q}_s}, ~~\mathcal{K}_s = X^{sta}W^{\mathcal{K}_s} \label{eq:stat1},~~\mathcal{V}_s = X^{sta}W^{\mathcal{V}_s}
\end{equation}
where $W^{\mathcal{Q}_s}$, $W^{\mathcal{K}_s}$ are the parameter matrices. Then, the statement attention scores are computed as:
\begin{equation}
\alpha_s = \frac{\mathcal{Q}_s\mathcal{K}_s^\mathrm{\emph{T}}}{\sqrt{d}} \label{eq:stat2}
\end{equation}
where $\alpha_s$ is the attention scores matrix. The value of the score $\alpha_s(i,j)$ denotes the relevant score between the $j$-th target token and the $i$-th source statement, and the scores with larger values demonstrate higher relevance. To make the model focus more on intent-relevant statements, we employ the top-$k$ selection strategy. Specifically, we reserve the $k$ largest scores of each row in $\alpha_s$ and set other scores in the row to negative infinity:
\begin{equation}
    \alpha_s(i,j) = \begin{cases}
    \alpha_s(i,j) &\alpha_s(i,j) \geq \alpha_s^k(i,*)  \\
    -\infty, &\alpha_s(i,j) < \alpha_s^k(i,*)
    \end{cases} \label{eq:stat3}
\end{equation}
where $k$ is a hyper-parameter, $\alpha_s^k(i,*)$ is the $k$-th largest score of row $i$. In this way, the most contributive statements for attention are reserved and other irrelevant information is filtered. 
We normalize the scores matrix with the Softmax function and obtain the statement-level attention $A_s$:
\begin{equation}
    A_{s} = {\rm Softmax}(\alpha_s) \label{eq:stat4}
\end{equation}
After the normalization, the attention weights between the target tokens and the unrelated source statements {will be approximately 0}.

\textbf{Token-level Attention Selection.}
Different from natural language, the programming language contains many tokens that are not associated with source code semantics \cite{DBLP:conf/sigsoft/NguyenNNN13}, such as the program separators (the period, the semicolon, parentheses, and braces).
Since such irrelevant tokens might frequently appear in the source code, they may be assigned high attention weights when the model attempts to get the information from the source code tokens. So this step aims to remove the distraction from those irrelevant tokens. 
Similar to the statement attention selection, we input the token-level representation $X^{tok}$, intent embedding $E$, and the query vector $Q^n_1$, and output the token-level attention $A_t^l$ for each statement $l$ using the equation (\ref{eq:stat1})-(\ref{eq:stat4}):

\textbf{Combining Attentions.} We combine the sentence-level attention $A_s$ and token-level attention $A_t$ to get the final selective attention matrix by conducting simple scalar element-wise multiplication:
\begin{equation}
A^l = A_s(l) \times A_t^l
\end{equation}
where $A^l$ is the final token-level attention of the $l$-th statement, $A_s(l)$ is the attention matrix for the $l$-th statement.
The intuition behind this is that when developers comment a code with a specific intent, they may first look for related statements in the whole code snippet and select the most important code token from these statements.
Finally, the output of ISA is computed as:
\begin{equation}
O^n_{\rm ISA} = AV_s
\end{equation}
where $A=[A^1, A^2, ..., A^L]$.

Following the distribution $A$, the attention can then become focused on the most contributive information.

\section{The Comment-Intent Labeling Tool}
\label{sec:tool}

Since training and evaluating our proposed approach require a large volume of labeled comment-intent data, we develop a COmment-INtent labeling tool, named {\ltool}, to support the automatic annotation of comment intents. This section introduces the design of {\ltool} and presents the analysis results of its effectiveness.

\subsection{The CodeBERT-based Comment-Intent Classifier}
Our comment-intent classifier utilizes the CodeBERT~\cite{feng2020codebert} as the backbone. CodeBERT is a powerful pre-trained language model built on top of the BERT-like~\cite{bert} architecture. It supports paired natural language and multi-lingual programming language data and has achieved great success in code search and code comment generation\cite{DBLP:conf/iclr/GuoRLFT0ZDSFTDC21,DBLP:conf/emnlp/0034WJH21}. 
We use the special separator tokens [CLS] and [SEP] to concatenate a comment and its corresponding code into a sequence and input it to the CodeBERT for embedding. 
The final token embedding of [CLS] is considered as the representation of the aggregated sequence. Then we feed the final embedding of [CLS] into a two-layer MLP followed by a softmax layer to obtain the probability of the comment intent. 
For training, we load the pre-trained parameters of CodeBERT\footnote{https://huggingface.co/microsoft/codebert-base} and fine-tune them with the cross entropy loss function on our annotated dataset that will be introduced later.

\begin{table}[thb!]
\centering
\caption{{Statistics of the manually-labeled intent
dataset}}
\label{table:dataset_classification}
\resizebox{0.6\columnwidth}{!}{
\begin{tabular}{ccc}
\toprule
Categoty       & Count & Proportion \\ 
\midrule
What           & 12,264 & 61.32\%    \\
Why            & 1,708  & 8.54\%    \\
How-to-use     & 573   & 2.87\%     \\
How-it-is-done & 2,933  & 14.67\%    \\ 
Property       & 2,270  & 11.35\%    \\ 
Others         & 252   & 1.26\%     \\
\bottomrule
\end{tabular}
}
\end{table}
\subsection{Effectiveness Evaluation}

\begin{table*}[hbtp!]
\centering
\caption{{The performance of intent classification for code comments}}
\label{table:classification}
\resizebox{\textwidth}{!}{

\begin{tabular}{c|ccc|ccc|ccc|ccc|ccc|ccc|ccc}
\hline
\multirow{2}{*}{Method} & \multicolumn{3}{c}{What} & \multicolumn{3}{c}{Why} & \multicolumn{3}{c}{How-to-use} & \multicolumn{3}{c}{How-it-is-done} & \multicolumn{3}{c}{Property} & \multicolumn{3}{c}{Others} & \multicolumn{3}{c}{Macro-Average} \\
\cline{2-22}
                        & P      & R      & F1     & P      & R      & F1    & P       & R      & F1     & P      & R      & F1     & P        & R       & F1      & P       & R       & F1     & P         & R         & F1        \\\hline
LGBM                    & 87.4   & 92.3   & 89.8   & 81.5   & 76.8   & 79.0  & 89.9    & 72.5   & 80.1   & 71.6   & 65.8   & 68.5   & 90.8     & 89.7    & 90.2    & 12.6    & 6.4     & 8.4    & 72.3      & 67.2      & 69.3      \\
RF                      & 83.9   & 94.3   & 88.8   & 84.0   & 73.2   & 78.2  & 92.3    & 70.3   & 79.6   & 75.8   & 54.5   & 63.4   & 90.9     & 89.3    & 90.0    & 39.9    & 29.7    & 30.3   & 77.8      & 68.5      & 71.7      \\
DT                      & 86.3   & 87.0   & 86.6   & 73.7   & 74.0   & 73.8  & 74.0    & 69.5   & 71.5   & 63.0   & 59.9   & 61.3   & 87.8     & 86.8    & 87.2    & 16.6    & 32.5    & 20.6   & 68.3      & 66.9      & 66.8      \\
CNN                     & 87.4   & 93.3   & 90.2   & 82.7   & 77.3   & 79.9  & 90.4    & 71.3   & 79.5   & 71.8   & 65.7   & 68.6   & 94.5     & 89.4    & 91.8    & 82.1    & 29.2    & 41.5   & 84.8      & 71.0      & 75.3      \\
BiLSTM                  & 88.4   & 89.0   & 88.6   & 78.5   & 74.6   & 76.4  & 81.0    & 62.8   & 70.4   & 60.5   & 68.8   & 64.2   & 96.1     & 88.2    & 92.0    & 10.0    & 0.9     & 1.7    & 69.1      & 64.1      & 65.5      \\
BiLSTM-Attn             & 87.3   & 91.5   & 89.3   & 79.5   & 75.2   & 77.2  & 80.2    & 68.1   & 73.3   & 68.2   & 67.3   & 67.7   & 96.1     & 88.6    & 92.1    & 0.0     & 0.0     & 0.0    & 68.5      & 65.1      & 66.6      \\\hline
\rowcolor{gray!25}{\ltool}                    & \textbf{92.5}   & \textbf{93.5}   & \textbf{93.0}   & \textbf{87.1}   & \textbf{86.1}   & \textbf{86.6}  & \textbf{93.0}    & \textbf{86.6}   & \textbf{89.6}   & \textbf{78.6}   & \textbf{79.2}   & \textbf{78.9}   & \textbf{96.4}     & \textbf{92.9}    & \textbf{94.6}    & \textbf{99.3}    & \textbf{90.7}    & \textbf{94.7}   & \textbf{91.1}      & \textbf{88.2}      & \textbf{89.6}     \\
\hline
\end{tabular}
}
\end{table*}
\subsubsection{Data Preparation} 
To train our comment-intent classifier, we randomly sample 20K code-comment pairs from two large-scale Java benchmark datasets (10K data for each), i.e., Funcom \cite{leclair2019neural} and TLC \cite{DBLP:conf/ijcai/HuLXLLJ18}, and invite five developers to manually classify the data into six intent categories.  

\textbf{Human Annotators.} We recruit five developers, including two senior researchers, one Ph.D. student, and two Master students, who are familiar with Java development and have at least three years of software development experience.

\textbf{Procedure.} To guarantee the accuracy of the labeling outcomes, we annotate each code-comment pair following a two-round process: 
First, two developers read the comment and the source code to decide its intent category. Each developer is assigned 8K code-comment pairs and annotates them independently. 
Second, after annotation, all five developers resolve conflicts via majority voting.
{The annotation process is labor-intensive. For each code-comment pair, developers need to read the source code and its one-sentence comment, and assign an intent category. One developer could label $\sim$ 64 pairs in an hour on average, and label $\sim$ 470 pairs per day. We spend 17 days for labeling the 8K pairs and $\sim$ 7.5 days for merging conflicts. Thus, the whole labeling process takes 24.5 days.}
The agreement between the two developers reaches 0.81 of Cohen's Kappa. The statistics of the final intent-labeled dataset are shown in Table~\ref{table:dataset_classification}.

\subsubsection{Baselines}

Chen et al. \cite{DBLP:journals/tosem/ChenXHLL21} experimented with four commonly used text classifiers (i.e., Light Gradient Boosting Machine (LGBM) \cite{DBLP:conf/nips/KeMFWCMYL17}, Random Forest (RF), Decision Tree (DT) \cite{DBLP:books/wa/BreimanFOS84}, and Bi-directional Long Short-Term Memory (BiLSTM) \cite{DBLP:journals/neco/HochreiterS97}) on classifying code comments into the five intents, and reported that the Random Forest classifier achieves the highest performance.
Following their work, we also include those commonly used text classifiers as our baselines.
Besides, we additionally add 
two neural-based classifiers (i.e., Convolutional Neural Network (CNN) \cite{DBLP:journals/pieee/LeCunBBH98} and BiLSTM+Attention\cite{DBLP:conf/acl/ZhouSTQLHX16}) into the comparison baselines.


\subsubsection{Evaluation Metrics}
Three commonly used metrics are used to evaluate the effectiveness for each category of comment intent, i.e., \textit{Precision}, \textit{Recall}, and \textit{F1}. 
Besides, as the comment-intent classification is a multi-class classification task, we also use the \textit{Macro-Precision}, \textit{Macro-Recall}, and \textit{Macro-F1} to evaluate the overall performance. 

\subsubsection{Results}
Table \ref{table:classification} demonstrates the performance of {\ltool}. Overall, {\ltool} outperforms other classifiers, which achieves 91.1\% of \textit{Macro-Precision}, 88.2\% of \textit{Macro-Recall}, and 89.6\% of \textit{Macro-F1} in 10-fold cross-validation. Compared with the best baseline classifier (CNN), {\ltool} improves the performance of \textit{Macro-Precision}, \textit{Macro-Recall}, and \textit{Macro-F1} by 7.43\%, 24.23\%, and 18.99\%, respectively. 
The results show that {\ltool} can achieve highly satisfactory performance on comment-intent classification, thereby enabling the automation of annotating intents for the code-comment dataset.
\section{Experimental Design}
\label{sec:exp}

\subsection{Dataset}
Since Funcom \cite{leclair2019neural} and TLC \cite{DBLP:conf/ijcai/HuLXLLJ18} are the most widely used benchmark datasets for code comment generation tasks~\cite{ahmad2020transformer,DBLP:journals/corr/abs-2111-08874,DBLP:journals/corr/abs-2104-09340,DBLP:conf/icse/ZhangW00020,DBLP:conf/icsm/LeClairBM21,DBLP:conf/emnlp/Shi0D0HZS21}, we select these two datasets to 
evaluate our approach in this study. 
Funcom contains 2.1M code-comment pairs from 29K Java projects, which were collected by Lopes \textit{et al.}\cite{Lopes} and cleaned by LeClair \textit{et al.}\cite{leclair2019neural}. TLC has 87,136 code-comment pairs collected from more than 9K Java GitHub repositories created from 2015 to 2016 with at least 20 stars. They first extracted Java methods and Javadocs, and treated the first sentence of the Javadoc as the ground-truth comment of the corresponding code. 
For the sake of fairness, we directly use the Funcom and TLC datasets open sourced by the previous work \cite{shi2022we}. They reported that many benchmark datasets have noisy data and provided a ``clean'' version of these datasets, which were cleaned by their automated cleaning tool CAT\footnote{https://github.com/BuiltOntheRock/FSE22\_BuiltOntheRock}.
After that, we use our trained comment-intent labeling tool {\ltool} to automatically annotate the comments in the two datasets with the corresponding intent categories. Since the \emph{others} comments are seen as unspecified or ambiguous comments, we exclude all data with the intent category of \emph{others}. In common with \cite{DBLP:conf/icse/ZhangW00020}, we further remove the exactly duplicated code-comment pairs in the test set for TLC dataset.
The statistics of the two preprocessed datasets are shown in Table \ref{table:dataset}. 
\begin{table}[tb!]
\centering
\vspace{-0.2cm}
\caption{Statistic of Funcom and TLC datasets}
\label{table:dataset}
\begin{tabular}{c|c|c}
\hline
Dataset                  & Funcom    & TLC  \\
\hline
Train                    & 1,178,923  & 53,528  \\
Valid               & 62,383   & 7,555  \\
Test                     & 69,259   & 4,985  \\
\hline
What & 762,884 & 36,604 \\
Why & 168,912 & 7,708 \\
How-to-use & 27,543 & 1,085 \\
How-it-is-done & 166,286 & 14,392 \\
Property & 184,940 & 6,279\\
\hline
\end{tabular}
\end{table}

\subsection{Evaluation Metrics}

We evaluate the performance of different approaches using common metrics including corpus BLEU~\cite{DBLP:conf/acl/PapineniRWZ02}, ROUGE-L~\cite{lin2004rouge}, and METEOR~\cite{DBLP:conf/acl/BanerjeeL05}.
\textbf{BLEU} is a standard evaluation metric in the code comment generation works. BLEU measures the $n$-gram precision by computing the overlap ratios of $n$-grams and applying a brevity penalty on short translation hypotheses.
\textbf{ROUGE-L} is defined as the length of the longest common subsequence between generated sentence and reference, and is based on recall scores.
\textbf{METEOR} is based on the harmonic mean of unigram precision and recall, with recall weighted higher than precision. 

{To ensure the consistency of metrics calculation, we calculate the values of the three metrics following the same scripts used in AST-Trans \cite{DBLP:conf/icse/TangSLGHZ022}.}
\begin{table*}[tp!]
\centering
\caption{{Performances of {\tool} and baselines on each intent category}}
\label{table:RQ1_RQ2_CR}
\resizebox{0.7\textwidth}{!}{
\renewcommand{\arraystretch}{1.2}
\begin{tabular}{c|c|c|ccc|ccc}
\hline
\multirow{2}{*}{Intent}   & \multicolumn{2}{c|}{\multirow{2}{*}{Method}}  & \multicolumn{3}{c}{Funcom}                       & \multicolumn{3}{c}{TLC}                          \\\cline{4-9}
                          & \multicolumn{2}{c|}{}                         & BLEU           & ROUGE-L        & METEOR         & BLEU           & ROUGE-L        & METEOR         \\
                          \hline
\multirow{9}{*}{What}     & \multirow{6}{*}{Baseline}     & Hybrid-DRL & 22.44          & 25.89         & 10.67      &19.69       &32.89          & 13.18           \\&  & CodeTrans      & 26.30       & 32.87        & 15.05        & 21.35       & 36.39         & 15.63          \\&  & Re$^2$Com      & 24.17       & 28.72         & 12.53        & 22.21          & 35.11          & 14.94         \\&  & Rencos       & 26.19          & 31.10          & 14.61          & 23.28          & 36.89          & 16.02          \\
                          &                               & EditSum      & 27.58          & 31.06          & 14.24          & 21.34          & 33.73          & 13.42          \\
                          &                               & AST-Trans    & 27.84          & 38.28          & 18.49          & 23.42          & 34.24          & 17.17          \\\cline{2-9}
                          & \multirow{3}{*}{\makecell{Our \\Approach}} & DOME         & \textbf{33.29} & \textbf{41.67} & \textbf{20.53} & \textbf{25.39} & \textbf{39.56} & \textbf{18.22} \\
                          &                               & DOME \textit{w/o ISA} & 32.01          & 38.74          & 18.65          & 24.37          & 37.93          & 16.51          \\
                          &                               & DOME \textit{w/o ER}  & 31.33          & 38.12          & 18.57          & 23.82          & 37.24          & 16.37          \\
                          \hline
\multirow{9}{*}{Why}      & \multirow{6}{*}{Baseline} & Hybrid-DRL      & 22.65       &27.61         & 11.14         & 16.47          & 28.61          & 11.56         \\&  & CodeTrans      & 26.52       & 35.04        & 15.88         & 17.58          & 32.18          & 13.96         \\&  & Re$^2$Com      & 24.39      & 30.75        & 13.22         & 18.99          & 31.45         & 13.09         \\&   & Rencos       & 24.81          & 30.12          & 14.21          & 20.55          & 32.99          & 14.54          \\
                          &                               & EditSum      & 27.17          & 30.98          & 14.66          & 18.42          & 29.85          & 11.81          \\
                          &                               & AST-Trans    & 25.96          & 35.74          & 17.77          & 19.31          & 29.37          & 14.94          \\\cline{2-9}
                          & \multirow{3}{*}{\makecell{Our \\Approach}} & DOME         & \textbf{33.07} & \textbf{42.31} & \textbf{20.56} & \textbf{21.97} & \textbf{35.31} & \textbf{15.77} \\
                          &                               & DOME \textit{w/o ISA} & 31.79          & 39.37          & 19.11          & 21.41          & 34.27          & 15.56          \\
                          &                               & DOME \textit{w/o ER}  & 31.13          & 38.78          & 18.99          & 19.60          & 32.16          & 12.71          \\
                          \hline
\multirow{9}{*}{How-to-use}    & \multirow{6}{*}{Baseline}  & Hybrid-DRL      & 21.16       & 24.75         & 10.04         & 10.25          & 19.78          & 7.45         \\&  & CodeTrans      & 25.79       & 32.91         & 15.24         & 13.50          & 24.26          & 10.28          \\&  & Re$^2$Com      & 23.17       & 27.79         & 12.18         & 14.18         & 23.45          & 10.09         \\&   & Rencos       & 25.54          & 27.74          & 13.34          & 14.62          & 22.61          & 10.21          \\
                          &                               & EditSum      & 25.25          & 27.25          & 12.79          & 14.00          & 21.51          & 9.07           \\
                          &                               & AST-Trans    & 24.93          & 30.90          & 15.09          & 13.24          & 18.20          & 9.16           \\\cline{2-9}
                          & \multirow{3}{*}{\makecell{Our \\Approach}} & DOME         & \textbf{31.63} & \textbf{39.31} & \textbf{19.34} & \textbf{17.16} & \textbf{26.11} & \textbf{12.36} \\
                          &                               & DOME \textit{w/o ISA} & 30.57          & 37.24          & 17.54          & 16.74          & 25.59          & 11.25          \\
                          &                               & DOME \textit{w/o ER}  & 30.42          & 37.15          & 17.37          & 15.33          & 24.96          & 10.94          \\
                          \hline
\multirow{9}{*}{How-it-is-done}     & \multirow{6}{*}{Baseline} & Hybrid-DRL      & 17.09       & 25.29         & 8.50         & 14.52          & 29.36          & 10.02          \\&  & CodeTrans      & 20.65       & 32.46         & 13.21         & 16.36          & 33.07          & 12.89          \\&  & Re$^2$Com      & 18.61      & 27.89         & 10.41         & 17.08          & 31.04          & 11.99          \\&    & Rencos       & 19.84          & 29.28          & 12.53          & 18.58          & 33.73          & 13.12          \\
                          &                               & EditSum      & 22.22          & 29.73          & 12.62          & 16.84          & 31.18          & 11.12          \\
                          &                               & AST-Trans    & 19.65          & 33.60          & 14.40          & 17.61          & 30.32          & 13.29          \\\cline{2-9}
                          & \multirow{3}{*}{\makecell{Our \\Approach}} & DOME         & \textbf{26.98} & \textbf{39.52} & \textbf{17.65} & \textbf{20.48} & \textbf{36.66} & \textbf{14.73} \\
                          &                               & DOME \textit{w/o ISA} & 26.03          & 38.19          & 18.20          & 19.50          & 36.57          & 13.10          \\
                          &                               & DOME \textit{w/o ER}  & 25.78          & 37.73          & 18.10          & 19.14          & 35.72          & 13.01          \\
                          \hline
\multirow{9}{*}{Property} & \multirow{6}{*}{Baseline} & Hybrid-DRL      & 23.30       & 34.84        & 15.09         & 21.53         & 39.37          & 17.17          \\&  & CodeTrans      & 27.00       & 41.95        & 19.53        & 22.70          & 42.63          & 19.33          \\&  & Re$^2$Com      & 24.85       & 37.77       & 17.08         & 23.37          & 40.79         & 18.92          \\&    & Rencos       & 25.57          & 35.60          & 16.39          & 23.82          & 38.85          & 17.76          \\
                          &                               & EditSum      & 26.79          & 36.09          & 16.60          & 22.35          & 37.74          & 16.33          \\
                          &                               & AST-Trans    & 28.29          & 43.54          & 20.73          & 23.54          & 36.56          & 18.47          \\\cline{2-9}
                          & \multirow{3}{*}{\makecell{Our \\Approach}} & DOME         & \textbf{34.18} & \textbf{49.43} & \textbf{24.32} & \textbf{26.01} & \textbf{45.73} & \textbf{21.29} \\
                          &                               & DOME \textit{w/o ISA} & 32.21          & 46.92          & 22.70          & 25.35          & 43.68          & 19.47          \\
                          &                               & DOME \textit{w/o ER}  & 31.80          & 46.09          & 21.59          & 25.01          & 42.85          & 19.16          \\
                          \hline\hline
\multirow{9}{*}{\textit{Average}}  & \multirow{6}{*}{Baseline}   & Hybrid-DRL      & 21.33       & 27.68        & 11.09         & 16.49          & 30.00            & 11.88          \\&  & CodeTrans      & 25.25      & 35.05        & 15.78         & 18.30          & 33.71          & 14.42          \\&  & Re$^2$Com      & 23.04       & 30.58         & 13.09        & 19.17          & 32.37          & 13.81          \\&     & Rencos       & 24.39          & 30.77          & 14.22          & 20.17          & 33.01          & 14.33          \\
                          &                               & EditSum      & 25.80          & 31.02          & 14.18          & 18.59          & 30.80          & 12.35          \\
                          &                               & AST-Trans    & 25.33          & 36.41          & 17.30          & 19.42          & 29.74          & 14.61          \\\cline{2-9}
                          & \multirow{3}{*}{\makecell{Our \\Approach}} & DOME         & \textbf{31.83} & \textbf{42.45} & \textbf{20.48} & \textbf{22.20} & \textbf{36.67} & \textbf{16.47} \\
                          &                               & DOME \textit{w/o ISA} & 30.52          & 40.09          & 19.24          & 21.47          & 35.61          & 15.18          \\
                          &                               & DOME \textit{w/o ER}  & 30.09          & 39.57          & 18.92          & 20.58          & 34.59          & 14.44 \\\hline        
\end{tabular}
}
\vspace{-0.4cm}
\end{table*}

\subsection{Implementation Details}
To train {\tool}, we first shuffle the training data and set the mini-batch size to 256 and 64 for Funcom and TLC datasets, respectively. For each batch, the code snippets and comments are padded with a special token [PAD] to the maximum length. Following previous studies \cite{DBLP:conf/icse/ZhangW00020,DBLP:conf/kbse/LiL000J21}, we limit the maximum length of the comment to 15 for Funcom and 30 for TLC. To save the computing resource, we limit the maximum vocabulary size to 50K and 30k for Funcom and TLC datasets. The out-of-vocabulary words are replaced by [UNK]. The word embedding size of both code and comment is set to 512, the dimension of intent embedding is set to 128. The $k$ is set to 10 for tokens and 5 for statements. We set the dimensions of hidden states to 512, the number of heads to 8, and the number of blocks to 6, respectively. We train our approach using the Adam \cite{DBLP:journals/corr/KingmaB14} optimizer with the learning rate 1e-4. To avoid the over-fitting problem, we apply dropout \cite{DBLP:journals/corr/abs-1207-0580} with 0.2. To reduce training time, we use the greedy search to generate comments at the training stage. During the prediction stage, we use the beam search~\cite{DBLP:conf/emnlp/WisemanR16} and set the beam size to 5. Our approach is implemented based on the Pytorch \cite{pytorch} framework. The experimental environment is a desktop computer equipped with an NVIDIA GeForce RTX 3060 GPU, intel core i5 CPU, and 12GB RAM, running on Ubuntu OS.

\section{Results}

\label{sec:result}
We address the following three research questions to evaluate the performance of {\tool}:

\textbf{RQ1 }: How does the {\tool} perform compared to the state-of-the-art comment generation baselines?

\textbf{RQ2}: How does each individual component in {\tool} contribute to the overall performance?

\textbf{RQ3}: What is the perceived quality of intent-aware comments generated by {\tool}?

\subsection{RQ1: Comparison with Baselines}

\subsubsection{Baselines}
We compare our approach with six state-of-the-art baselines on the comment generation task. All baselines adopted the hyper-parameter settings reported in the original paper. {To ensure the proper implementation, we reproduced the six baselines on the same datasets provided in their paper, and have verified they achieved comparable results as the original paper reported.}

\begin{itemize}[leftmargin=*]
\item{\textbf{Hybrid-DRL}}\cite{DBLP:conf/kbse/WanZYXY0Y18} is a novel reinforcement learning comment generation framework that incorporates an Abstract Syntax Tree (AST) structure as well as sequential content of code snippets into a deep actor-critic network.
\item{\textbf{CodeTrans}}\cite{ahmad2020transformer} is a transformer-based approach that uses relative distances instead of absolute positions in the attention computation and applies copy mechanism to copy rare tokens from the input source code.
while only relying on language-agnostic features
\item{\textbf{Re$^2$Com}}\cite{DBLP:conf/kbse/WeiLLXJ20} is an exemplar-based comment generation approach that leverages the advantages of three types of methods based on neural networks, templates, and IR to improve the performance. 
\item{\textbf{Rencos}}\cite{DBLP:conf/icse/ZhangW00020} is a hybrid approach that combines the advantages of both IR-based and NMT-based techniques. 
Given a code snippet for testing, Rencos retrieves its two most similar code snippets in the training set from the aspects of syntax and semantics, and input the three code into the encoder-decoder model to predict the comment.
\item{\textbf{EditSum}}\cite{DBLP:conf/kbse/LiL000J21} is a retrieve-and-edit framework for code comment generation. Given a code snippet, EditSum first retrieves its most similar code snippet, and treats the corresponding comment as a prototype. Then, it combines the pattern in the prototype and semantic information of the input code to generate the target comment.
\item{\textbf{AST-Trans}}
 \cite{DBLP:conf/icse/TangSLGHZ022} is the most recent study for comment generation using a novel Transformer-based model. AST-Trans leverages tree-structured attention to dynamically assign weights to related nodes, while considering the ancestor-descendant and sibling relationships of AST. This structure information is incorporated into the model to generate the target comments.
 \end{itemize}
 
\subsubsection{Setting}
As aforementioned, this study is the first work to generate various comments for one code snippet. The existing comment generation approaches are trained to learn a one-to-one mapping, and cannot generate various comments given different intents. Thus, to compare the effectiveness of comment generation with different intents, we first divide the test dataset into five groups according to the automated annotation of comment intents. Then we train all baselines and our approach on the same training set, test them on each intent group of the test set, and obtain their performance (i.e., BLEU, ROUGE, and METEOR) on each intent category. In order to facilitate the overall comparison, we also calculate the average performance of all approaches on the five intent categories.

 
\subsubsection{Results}
Table \ref{table:RQ1_RQ2_CR} shows the comparison results between the performance of {\tool} and other baselines, and the best performance is highlighted in bold. 
Overall, our approach achieves the best performance on all evaluation metrics. 
On Funcom dataset, {{\tool} achieves 31.83, 42.45, and 20.48 points on BLEU, ROUGE-L, and METEOR.} Compared with the best baseline (AST-Trans), {\tool} improves the performance of BLEU, ROUGE-L, and METEOR by 25.66\%, 16.59\%, and 18.38\%, respectively. 
On TLC dataset, {{\tool} achieves 22.20, 36.67, and 16.47 on BLEU, ROUGE-L, and METEOR.} 
Compared with the best baseline (Rencos), {\tool} also achieves 10.06\%, 11.09\%, and 14.93\% improvements on the three metrics.
For each intent category, the performance of {\tool}  outperforms all the other baselines. It is mainly because {\tool} can effectively utilize the intent information to guide the model to generate relevant and fluent comments.

\mybox{
\textbf{Answering RQ1:} For each intent category,  {\tool} outperforms the state-of-the-art baselines in terms of three metrics on both two datasets. Overall, compared to the best baselines, {\tool} improves the performance of BLEU, ROUGE-L, and METEOR by 25.66\%, 16.59\%, and 18.38\% on Funcom dataset, by 10.06\%, 11.09\%, and 14.93\% on TLC dataset, respectively.}

\subsection{RQ2: Component Analysis}

\subsubsection{Variants}
To evaluate the contribution of core components, we obtain two variants: (1) \textbf{{\tool} w/o ISA}, which replaces the intent-guided selective attention with the vanilla cross attention to generate comments. (2) \textbf{{\tool} w/o ER}, which removes the exemplar retriever and only uses the encoder-decoder framework to generate comments.
{We train the two variants with the same experimental setup as {\tool} and evaluate their performance on the test sets of Funcom and TLC, respectively.}

\begin{figure*}[t]
\centering
\includegraphics[width=0.95\textwidth]{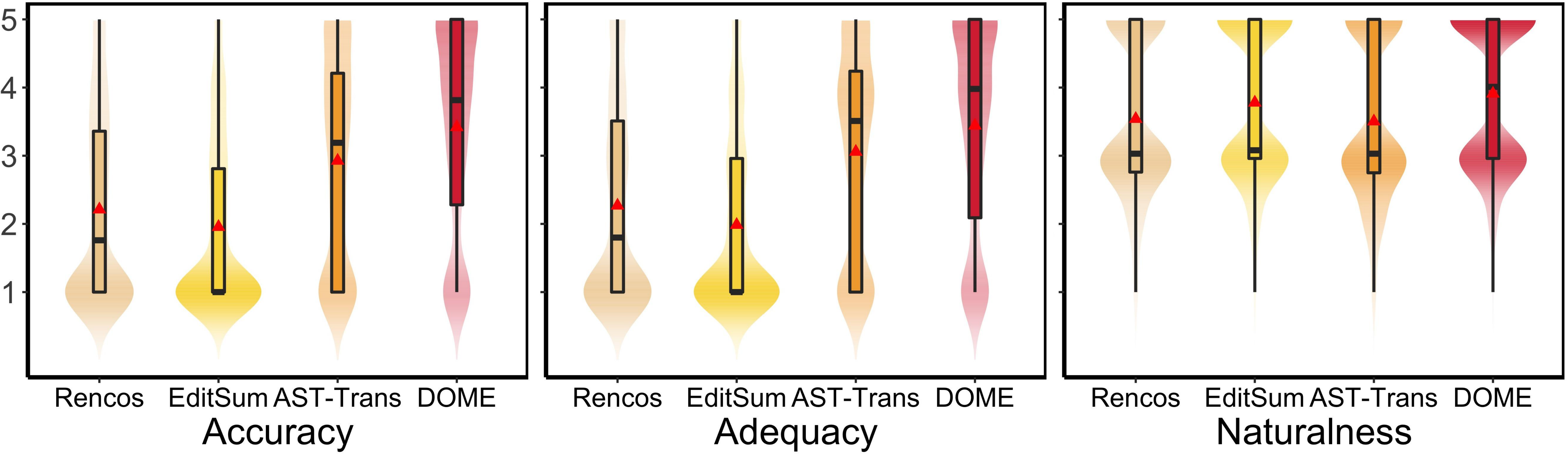}
\caption{{The results of human evaluation.}}
\vspace{-0.5cm}
\label{fig:RQ3}
\end{figure*}

\subsubsection{Results}
Table \ref{table:RQ1_RQ2_CR} presents the performances of {\tool} and its two variants. 
We can see that, removing the two components makes the performance degrade substantially. Specifically, when comparing {\tool} and {\tool} w/o ISA, removing the selective attention will lead to a dramatic decrease in the average BLEU (by 3.70\%), ROUGE-L (by 4.23\%), and METEOR (by 6.94\%) across both datasets. 
{When comparing {\tool} and {\tool} w/o ER,
we find that removing the exemplar retriever will lead to the performance decline in the average BLEU (by 6.38\%), ROUGE-L (by 6.23\%), and METEOR (by 9.97\%).}

\mybox{
\textbf{Answering RQ2:} Both the ISA and the ER components have positive contributions to the performance of {\tool}.}


\subsection{RQ3: Human Evaluation}
\label{RQ3}
Although the evaluation metrics (i.e., BLEU, ROUGE-L, and METEOR) can measure the lexical gap between the generated comments and the references,
they can hardly reflect the semantic gap. 
Therefore, we perform a human evaluation to further assess the quality of comments generated by different approaches.

\subsubsection{Procedure}
We crawl the 10 most-star Java projects on Github, and use the automated preprocessing tool CAT \cite{shi2022we} to preprocess these data. There are 100 code snippets that are randomly selected from the preprocessed data. 
{For each code snippet, we first let each participant select one or more types of intents they would like to write the comment. 
Then, we generate the comments on the developers' demand by using {\tool} as well as the three best-performing baselines (i.e., Rencos, EditSum, and AST-Trans). 
In total, we obtain 400 generated comments as our evaluation subjects.}


We recruited six participants, including three Ph.D. students, one Master student, and two senior researchers. They all have at least three years of Java development experience, and four of them have more than six years of development experience.
{Note that, all the participants are not co-authors of this paper. We signed agreements with all the participants, which explicitly required them to annotate or evaluate objectively.}
{For each participant, we assign 33-34 code snippets. Each code snippet has four comments, and is evaluated by two participants. To ensure fairness, the participants are not aware of where the comments are generated from.}
Each participant is asked to rate each comment from the three aspects: (1) \textbf{Accuracy} reflects the accuracy of generated comment from the perspective of whether its content is consistent with the code,
(2) \textbf{Adequacy} refers to whether the generated comments are missing information in the source code, and (3) \textbf{Naturalness} reflects the fluency of generated text from the perspective of grammar. All three scores are integers, ranging from 1 to 5. {Higher score indicates better performance.}

\subsubsection{Results}

Figure \ref{fig:RQ3} exhibits the results of human evaluation by showing the violin plots depicting the accuracy, adequacy, and naturalness of comments generated by different models.
Overall, the quality of comments generated by {\tool} is better than all baselines in three aspects. 
The average score for accuracy, adequacy, and naturalness of comments generated by our approach are 3.41, 3.44, and 3.91, respectively. Compared with the best baseline results, {\tool} achieves 14.43\%, 11.17\%, and 3.38\% improvements in accuracy, adequacy, and naturalness.
{The results indicate that the comments generated by {\tool} tend to be more informative, accurate, and fluent than other baselines.}

\mybox{
\textbf{Answering RQ3:} In human evaluation, compared to the baselines, {\tool} achieves the highest scores on accuracy, adequacy, and naturalness, respectively.}  
\section{Discussion}
\label{sec:discussion}
\subsection{Qualitative Analysis and Attention Visualization}

\begin{figure*}[t]
\vspace{0.6cm}
\centering
\includegraphics[width=0.95\textwidth]{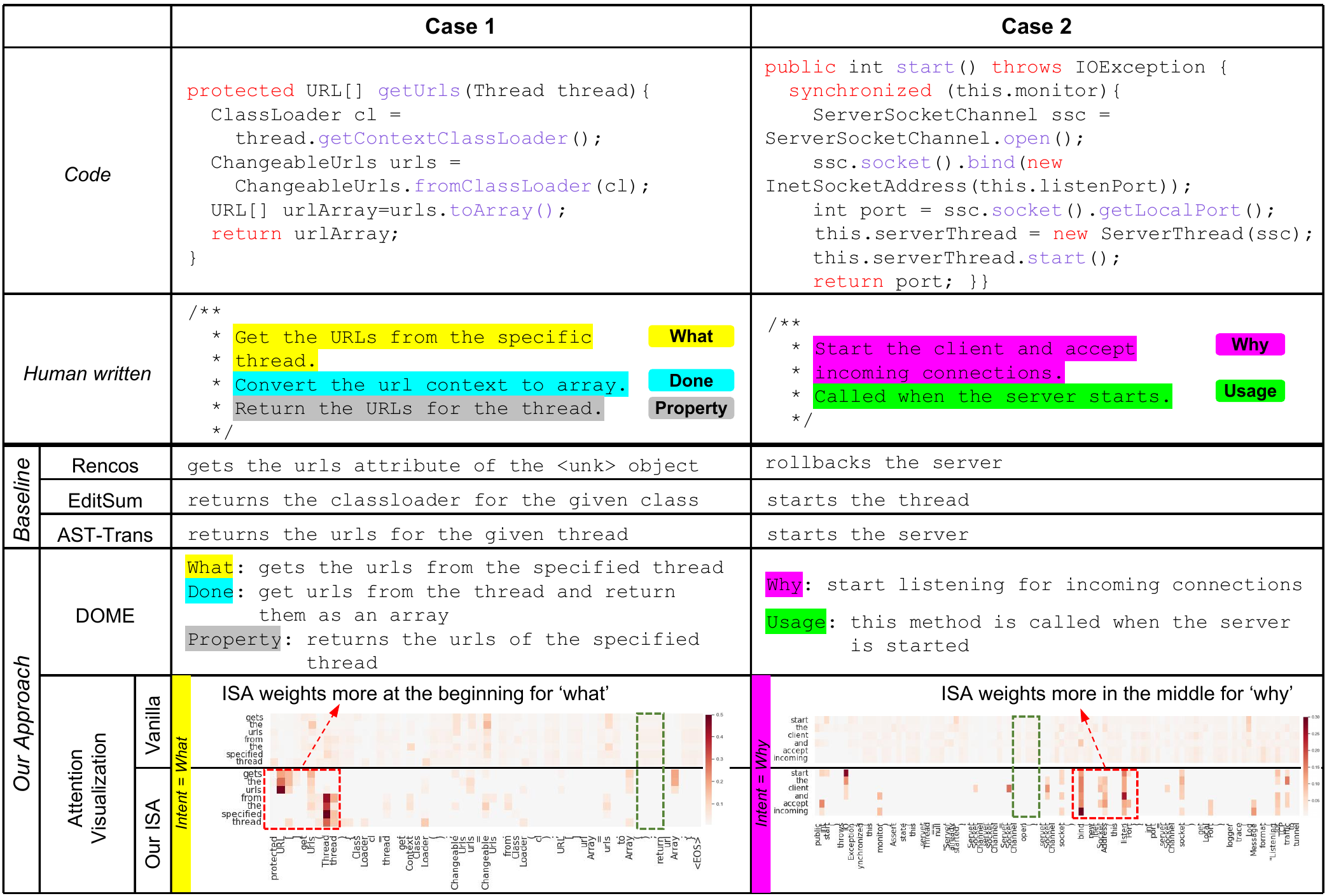}
\caption{{Examples of comments generated by each model and attention visualization of {\tool}.}}
\vspace{-0.3cm}
\label{fig:example}
\end{figure*}

For qualitative analysis of our approach, we present two cases generated by the three best-performing baselines together with {\tool}. The cases are selected from the real-world Java projects which we introduced in Section \ref{RQ3}. 

\textbf{Comment Analysis.}
As shown in Figure \ref{fig:example}, Given a code snippet, each SOTA baseline can only generate a short comment with a single intent, while our method can produce comments with multiple intents.
In Case 1, the comment generated by Rencos summarizes the functionality of the method \textit{getURLs}. However, it predicts a wrong word ``attribute'' and an out-of-vocabulary word that has been replaced by [UNK]. In contrast, the \emph{what} comment generated by {\tool} is exactly the same as the human-written comment. Besides, Compared with the comments generated by Editsum and AST-Trans which describe the property of the code, the \emph{property} comment generated by {\tool} is more accurate and fluent.
In Case 2, we can see that, all three comments generated by baselines are short and less informative. While the \emph{why} and \emph{usage} comments generated by {\tool} have a high semantic similarity with the human-written comment.
The two cases indicate that our approach can generate multiple accurate and fluent comments that reflect different intents appropriately.
Thereby {\tool} could better satisfy the scenarios of real-world comment practice. 

\textbf{Attention Visualization.}
{We further visualize the vanilla cross attention and ISA of the two cases in Figure \ref{fig:example}. 
Taking the \emph{what} comment ``gets the urls from the specified thread'' in Case 1 as an example, 
we can notice that the distribution of the vanilla cross attention is fairly dispersed, showing that it cannot concentrate on the important code tokens. In contrast, ISA concentrates on the beginning part of method \textit{getURLs}, which contains many important tokens, such as ``get'', ``URL'', and ``thread''. Besides, the vanilla attention assigns many weights to the program separators (in the green box), which may introduce noise into the model. While ISA removes the distraction from irrelevant tokens based on the top-$k$ selection. 
Taking the \emph{why} comment ``start listening for incoming connections'' in Case 2 as another example, ISA pays more attention to the middle part of the method \textit{start()} and less attention to irrelevant tokens. The visualization of the two attention variants shows that ISA can enable the attention more concentrated on the most contributive tokens or statements in the source code based on the given intent.
}

\subsection{Threats to Validity}

The first threat to validity is the assumption that {\tool} can retrieve a similar comment in the retrieval corpus. It is limited by two aspects: (1) The retrieval model is not powerful enough to find similar comments (2) Very similar comments do not exist in the retrieval corpus. To mitigate this threat, first, {\tool} employs the SOTA retrieve model DPR \cite{DBLP:conf/emnlp/KarpukhinOMLWEC20} that has a better performance than the traditional term-based methods.
Second, {\tool} introduces a gated fusion layer to dynamically decide whether to use the semantic features from the retrieved exemplar. Thus, even though the dissimilar exemplar is retrieved, DOME still can guarantee its performance is not affected.

The second threat to validity is the datasets we use. 
We only evaluate {\tool} on two Java datasets. Although Java may not be representative of all programming languages, the experimental datasets are large and safe enough to show the effectiveness of our model. Furthermore, {\tool} uses language-agnostic features that can be easily extracted from any programming language. Therefore, we believe that our approach has good generalizability, and can perform well on the datasets of other programming languages as well, such as Python and C\#.

The third threat relates to the suitability of evaluation metrics. First, recent researchers have raised concern over the use of BLEU, warning the community that the way BLEU is used and interpreted can significantly affect its reliability. To mitigate that threat, we also adopt other metrics, i.e., ROUGE-L, and METEOR when evaluating performance.
Besides, we perform a human evaluation to further assess the quality of comments generated by {\tool} in terms of accuracy, adequacy, and naturalness, and whether {\tool} can meet the needs of developers in real-world usage scenarios.

\section{Related Work}
\label{sec:related}
\subsection{Automatic Comment Generation}
The automatic comment generation task now is a rapidly-growing research topic in the community of software engineering and natural language processing. 
Early studies typically utilize template-based approaches~\cite{DBLP:conf/kbse/SridharaHMPV10,DBLP:conf/iwpc/MorenoASMPV13,DBLP:journals/tse/McBurneyM16} and information retrieval (IR) based approaches~\cite{DBLP:conf/icse/HaiducAM10,DBLP:conf/wcre/HaiducAMM10,DBLP:conf/iwpc/EddyRKC13,DBLP:conf/kbse/WongYT13,DBLP:conf/wcre/WongLT15,DBLP:journals/jasis/DeerwesterDLFH90,DBLP:journals/cacm/SaltonWY75,liu2018neural} to generate comments. 
Recently, many learning-based methods have been proposed, which train the neural models from a large-scale code-comment corpus to automatically generate comments \cite{DBLP:conf/acl/IyerKCZ16,DBLP:conf/ijcai/HuLXLLJ18,DBLP:conf/kbse/WanZYXY0Y18,DBLP:conf/kbse/WeiLLXJ20,DBLP:conf/iwpc/HuLXLJ18,DBLP:conf/icse/ZhangW00020,DBLP:conf/kbse/LiL000J21, liu2021haconvgnn,9031440,DBLP:journals/tosem/WangXLHWG21,DBLP:conf/icsm/LeClairBM21,DBLP:conf/icse/TangSLGHZ022, DBLP:conf/kbse/MuC0WW22}.
Iyer \textit{et al.} \cite{DBLP:conf/acl/IyerKCZ16} first treated the comment generation task as an end-to-end translation problem and introduced NMT techniques into code comment generation. 
Zhang \textit{et al.} \cite{DBLP:conf/icse/ZhangW00020} proposed a seq2seq approach that retrieved two similar code snippets for a given code to improve the quality of the generated comment.
Li \textit{et al.} \cite{DBLP:conf/kbse/LiL000J21} treated the comment of the similar code retrieved from a parallel corpus as a prototype. They proposed a seq2seq network to update the prototype and generate comments. 
Further, Tang \textit{et al.} \cite{DBLP:conf/icse/TangSLGHZ022} proposed AST-Trans which exploits two types of node relationships in the AST: ancestor-descendant and sibling relationships. AST-Trans adopts the tree-structure attention to learn this structure information, thereby generating high-quality comments.

Although existing research has achieved promising results in comment generation task, they only focus on creating a general description of functionality for a given code snippet without considering developer intentions, which may have limitations in practical usage. Our work aims to bridge the gap and defines a developer-intent driven comment generation task that can generate intent-aware comments for the same source code with different intents.

\subsection{Controllable Text Generation}
On the basis of traditional text generation, controllable text generation makes the output text more personalized or standardized by introducing the control element, such as text style or key information. 
This technology has broad application prospects in NLP, such as attribute-based generation, Data Augmentation, and format control\cite{zhang2022survey}.
Hu \textit{et al.} \cite{pmlr-v70-hu17e} proposed a framework based on VAEs, which can generate sentences according to the language attributes, such as sentiment and tenses.
Zhou \textit{et al.} \cite{zhou2018emotional} used a trained emotion classifier to label the data. According to different emotion categories, the model will generate responses with different emotions.
Xu \textit{et al.} \cite{xu2020megatron} proposed a framework composed of a keyword predictor, knowledge retriever, and knowledge ranker, etc., which combines with an external knowledge database to control the story generation.
Keskar \textit{et al.} \cite{DBLP:journals/corr/abs-1909-05858} released a large conditional transformer language model named CTRL, whose output text is controlled by the given style, content, and task-specific behavior.

{Our study is different from the previous work as we focus on leveraging controllable text generation techniques to improve the comment generation task. To the best of our knowledge, this is the first work that treats the comment generation task as a one-to-many generation task, and utilizes the intent to control the content and style of the generated comments.}
\section{Conclusion}
In this work, we focus on solving the developer-intent driven comment generation task, which requires the model to generate intent-aware comments given the same source code and different intents of developers. To solve this challenging task, we propose a novel method, named DOME, which first incorporates developer intents in comment generation and can create a comment that is coherent with the given intent. Specifically, DOME first utilizes the DPR model to retrieve the most similar comment as the exemplar. Then, it inputs the source code and the retrieved exemplar into two encoders to encode them into representation sequences respectively. Next, it utilizes intent-guided selective attention to explicitly select intent-relevant information, and removes irrelevant noise from the source code. Finally, the semantic features of the code and examples are fused to generate the final comment.
Furthermore, since training and evaluating DOME require a large volume of labeled comment-intent data, we developed an automated comment-intent labeling tool {\ltool} that can be used to construct high-quality intent-annotated code comment datasets. We evaluate DOME on two real-world Java datasets, and the experimental results show that our approach outperforms the state-of-the-art baselines. 
A human evaluation also confirms the significant potential of applying {\tool} in practical usage, enabling developers to comment code effectively according to their own needs.
\section*{ACKNOWLEDGMENTS}
We sincerely appreciate anonymous reviewers for their constructive and insightful suggestions for improving this manuscript. 
This work is supported by the National Natural Science Foundation of China Grant No.62272445, No.62232016, and No.62072442, and Youth Innovation Promotion Association Chinese Academy of Sciences.

\bibliographystyle{IEEEtran}
\bibliography{ref}

\begin{thebibliography}{10}
\providecommand{\url}[1]{#1}
\csname url@samestyle\endcsname
\providecommand{\newblock}{\relax}
\providecommand{\bibinfo}[2]{#2}
\providecommand{\BIBentrySTDinterwordspacing}{\spaceskip=0pt\relax}
\providecommand{\BIBentryALTinterwordstretchfactor}{4}
\providecommand{\BIBentryALTinterwordspacing}{\spaceskip=\fontdimen2\font plus
\BIBentryALTinterwordstretchfactor\fontdimen3\font minus
  \fontdimen4\font\relax}
\providecommand{\BIBforeignlanguage}[2]{{%
\expandafter\ifx\csname l@#1\endcsname\relax
\typeout{** WARNING: IEEEtran.bst: No hyphenation pattern has been}%
\typeout{** loaded for the language `#1'. Using the pattern for}%
\typeout{** the default language instead.}%
\else
\language=\csname l@#1\endcsname
\fi
#2}}
\providecommand{\BIBdecl}{\relax}
\BIBdecl

\bibitem{DBLP:journals/tosem/ChenXHLL21}
Q.~Chen, X.~Xia, H.~Hu, D.~Lo, and S.~Li, ``{Why My Code Summarization Model
  Does Not Work: Code Comment Improvement with Category Prediction},''
  \emph{{ACM} Trans. Softw. Eng. Methodol.}, vol.~30, no.~2, pp. 25:1--25:29,
  2021.

\bibitem{DBLP:conf/icse/ZhaiXSTPMXZTZ20}
J.~Zhai, X.~Xu, Y.~Shi, G.~Tao, M.~Pan, S.~Ma, L.~Xu, W.~Zhang, L.~Tan, and
  X.~Zhang, ``{{CPC:} Automatically Classifying and Propagating Natural
  Language Comments via Program Analysis},'' in \emph{{ICSE} '20: 42nd
  International Conference on Software Engineering}.\hskip 1em plus 0.5em minus
  0.4em\relax {ACM}, 2020, pp. 1359--1371.

\bibitem{DBLP:conf/emnlp/KarpukhinOMLWEC20}
V.~Karpukhin, B.~Oguz, S.~Min, P.~S.~H. Lewis, L.~Wu, S.~Edunov, D.~Chen, and
  W.~Yih, ``{Dense Passage Retrieval for Open-Domain Question Answering},'' in
  \emph{Proceedings of the 2020 Conference on Empirical Methods in Natural
  Language Processing, {EMNLP} 2020}.\hskip 1em plus 0.5em minus 0.4em\relax
  Association for Computational Linguistics, 2020, pp. 6769--6781.

\bibitem{leclair2019neural}
A.~LeClair, S.~Jiang, and C.~McMillan, ``{A Neural Model for Generating Natural
  Language Summaries of Program Subroutines},'' in \emph{2019 IEEE/ACM 41st
  International Conference on Software Engineering (ICSE)}, pp. 795--806.

\bibitem{DBLP:conf/ijcai/HuLXLLJ18}
X.~Hu, G.~Li, X.~Xia, D.~Lo, S.~Lu, and Z.~Jin, ``{Summarizing Source Code with
  Transferred {API} Knowledge},'' in \emph{Proceedings of the Twenty-Seventh
  International Joint Conference on Artificial Intelligence, {IJCAI} 2018},
  2018, pp. 2269--2275.

\bibitem{website}
``{Project Website},'' \url{https://github.com/ICSE-DOME/DOME}, 2022.

\bibitem{DBLP:conf/nips/VaswaniSPUJGKP17}
A.~Vaswani, N.~Shazeer, N.~Parmar, J.~Uszkoreit, L.~Jones, A.~N. Gomez,
  L.~Kaiser, and I.~Polosukhin, ``{Attention is All You Need},'' in
  \emph{Advances in Neural Information Processing Systems 30: Annual Conference
  on Neural Information Processing Systems 2017}, 2017, pp. 5998--6008.

\bibitem{DBLP:conf/acl/WangLXZLWC19}
Q.~Wang, B.~Li, T.~Xiao, J.~Zhu, C.~Li, D.~F. Wong, and L.~S. Chao, ``{Learning
  Deep Transformer Models for Machine Translation},'' in \emph{Proceedings of
  the 57th Conference of the Association for Computational Linguistics, {ACL}
  2019}.\hskip 1em plus 0.5em minus 0.4em\relax Association for Computational
  Linguistics, 2019, pp. 1810--1822.

\bibitem{DBLP:conf/emnlp/LiuL19}
Y.~Liu and M.~Lapata, ``{Text Summarization with Pretrained Encoders},'' in
  \emph{Proceedings of the 2019 Conference on Empirical Methods in Natural
  Language Processing and the 9th International Joint Conference on Natural
  Language Processing, {EMNLP-IJCNLP} 2019}.\hskip 1em plus 0.5em minus
  0.4em\relax Association for Computational Linguistics, 2019, pp. 3728--3738.

\bibitem{DBLP:journals/corr/BaKH16}
L.~J. Ba, J.~R. Kiros, and G.~E. Hinton, ``{Layer Normalization},''
  \emph{CoRR}, vol. abs/1607.06450, 2016.

\bibitem{DBLP:conf/icse/ZhangW00020}
J.~Zhang, X.~Wang, H.~Zhang, H.~Sun, and X.~Liu, ``{Retrieval-based Neural
  Source Code Summarization},'' in \emph{{ICSE} '20: 42nd International
  Conference on Software Engineering}.\hskip 1em plus 0.5em minus 0.4em\relax
  {ACM}, 2020, pp. 1385--1397.

\bibitem{DBLP:conf/kbse/LiL000J21}
J.~Li, Y.~Li, G.~Li, X.~Hu, X.~Xia, and Z.~Jin, ``{EditSum: {A}
  Retrieve-and-Edit Framework for Source Code Summarization},'' in \emph{36th
  {IEEE/ACM} International Conference on Automated Software Engineering, {ASE}
  2021}.\hskip 1em plus 0.5em minus 0.4em\relax {IEEE}, 2021, pp. 155--166.

\bibitem{DBLP:conf/kbse/WeiLLXJ20}
B.~Wei, Y.~Li, G.~Li, X.~Xia, and Z.~Jin, ``{Retrieve and Refine:
  Exemplar-based Neural Comment Generation},'' in \emph{35th {IEEE/ACM}
  International Conference on Automated Software Engineering, {ASE} 2020},
  2020, pp. 349--360.

\bibitem{DBLP:journals/jd/Jones04}
K.~S. Jones, ``{A Statistical Interpretation of Term Specificity and Its
  Application in Retrieval},'' \emph{J. Documentation}, vol.~60, no.~5, pp.
  493--502, 1972.

\bibitem{DBLP:journals/ftir/RobertsonZ09}
S.~E. Robertson and H.~Zaragoza, ``{The Probabilistic Relevance Framework:
  {BM25} and Beyond},'' \emph{Found. Trends Inf. Retr.}, vol.~3, no.~4, pp.
  333--389, 2009.

\bibitem{DBLP:conf/iclr/LiuCXS021}
S.~Liu, Y.~Chen, X.~Xie, J.~K. Siow, and Y.~Liu, ``{Retrieval-Augmented
  Generation for Code Summarization via Hybrid {GNN}},'' in \emph{9th
  International Conference on Learning Representations, {ICLR} 2021}.\hskip 1em
  plus 0.5em minus 0.4em\relax OpenReview.net, 2021.

\bibitem{DBLP:conf/emnlp/ParvezACRC21}
M.~R. Parvez, W.~U. Ahmad, S.~Chakraborty, B.~Ray, and K.~Chang, ``{Retrieval
  Augmented Code Generation and Summarization},'' in \emph{Findings of the
  Association for Computational Linguistics: {EMNLP} 2021, Virtual Event /
  Punta Cana, Dominican Republic}.\hskip 1em plus 0.5em minus 0.4em\relax
  Association for Computational Linguistics, 2021, pp. 2719--2734.

\bibitem{DBLP:conf/icse/ZhangWZ0WL19}
J.~Zhang, X.~Wang, H.~Zhang, H.~Sun, K.~Wang, and X.~Liu, ``{A Novel Neural
  Source Code Representation Based on Abstract Syntax Tree},'' in
  \emph{Proceedings of the 41st International Conference on Software
  Engineering, {ICSE} 2019}.\hskip 1em plus 0.5em minus 0.4em\relax {IEEE} /
  {ACM}, 2019, pp. 783--794.

\bibitem{DBLP:journals/neco/HochreiterS97}
S.~Hochreiter and J.~Schmidhuber, ``{Long Short-Term Memory},'' \emph{Neural
  Comput.}, vol.~9, no.~8, pp. 1735--1780, 1997.

\bibitem{DBLP:conf/sigsoft/NguyenNNN13}
T.~T. Nguyen, A.~T. Nguyen, H.~A. Nguyen, and T.~N. Nguyen, ``{A Statistical
  Semantic Language Model for Source Code},'' in \emph{Joint Meeting of the
  European Software Engineering Conference and the {ACM} {SIGSOFT} Symposium on
  the Foundations of Software Engineering, ESEC/FSE'13}.\hskip 1em plus 0.5em
  minus 0.4em\relax {ACM}, 2013, pp. 532--542.

\bibitem{feng2020codebert}
Z.~Feng, D.~Guo, D.~Tang, N.~Duan, X.~Feng, M.~Gong, L.~Shou, B.~Qin, T.~Liu,
  D.~Jiang, and M.~Zhou, ``{CodeBERT: A Pre-Trained Model for Programming and
  Natural Languages},'' 2020.

\bibitem{bert}
J.~Devlin, M.~Chang, K.~Lee, and K.~Toutanova, ``{{BERT:} Pre-training of Deep
  Bidirectional Transformers for Language Understanding},'' in
  \emph{Proceedings of the 2019 Conference of the North American Chapter of the
  Association for Computational Linguistics: Human Language Technologies,
  {NAACL-HLT} 2019}.\hskip 1em plus 0.5em minus 0.4em\relax Association for
  Computational Linguistics, 2019, pp. 4171--4186.

\bibitem{DBLP:conf/iclr/GuoRLFT0ZDSFTDC21}
D.~Guo, S.~Ren, S.~Lu, Z.~Feng, D.~Tang, S.~Liu, L.~Zhou, N.~Duan,
  A.~Svyatkovskiy, S.~Fu, M.~Tufano, S.~K. Deng, C.~B. Clement, D.~Drain,
  N.~Sundaresan, J.~Yin, D.~Jiang, and M.~Zhou, ``{GraphCodeBERT: Pre-training
  Code Representations with Data Flow},'' in \emph{9th International Conference
  on Learning Representations, {ICLR} 2021}.\hskip 1em plus 0.5em minus
  0.4em\relax OpenReview.net, 2021.

\bibitem{DBLP:conf/emnlp/0034WJH21}
Y.~Wang, W.~Wang, S.~R. Joty, and S.~C.~H. Hoi, ``{CodeT5: Identifier-aware
  Unified Pre-trained Encoder-Decoder Models for Code Understanding and
  Generation},'' in \emph{Proceedings of the 2021 Conference on Empirical
  Methods in Natural Language Processing, {EMNLP} 2021}.\hskip 1em plus 0.5em
  minus 0.4em\relax Association for Computational Linguistics, 2021, pp.
  8696--8708.

\bibitem{DBLP:conf/nips/KeMFWCMYL17}
G.~Ke, Q.~Meng, T.~Finley, T.~Wang, W.~Chen, W.~Ma, Q.~Ye, and T.~Liu,
  ``{LightGBM: {A} Highly Efficient Gradient Boosting Decision Tree},'' in
  \emph{Advances in Neural Information Processing Systems 30: Annual Conference
  on Neural Information Processing Systems 2017}, 2017, pp. 3146--3154.

\bibitem{DBLP:books/wa/BreimanFOS84}
L.~Breiman, J.~H. Friedman, R.~A. Olshen, and C.~J. Stone,
  \emph{{Classification and Regression Trees}}.\hskip 1em plus 0.5em minus
  0.4em\relax Wadsworth, 1984.

\bibitem{DBLP:journals/pieee/LeCunBBH98}
Y.~LeCun, L.~Bottou, Y.~Bengio, and P.~Haffner, ``{Gradient-based Learning
  Applied to Document Recognition},'' \emph{Proc. {IEEE}}, vol.~86, no.~11, pp.
  2278--2324, 1998.

\bibitem{DBLP:conf/acl/ZhouSTQLHX16}
P.~Zhou, W.~Shi, J.~Tian, Z.~Qi, B.~Li, H.~Hao, and B.~Xu, ``{Attention-Based
  Bidirectional Long Short-Term Memory Networks for Relation Classification},''
  in \emph{Proceedings of the 54th Annual Meeting of the Association for
  Computational Linguistics, {ACL} 2016}.\hskip 1em plus 0.5em minus
  0.4em\relax The Association for Computer Linguistics, 2016.

\bibitem{ahmad2020transformer}
W.~U. Ahmad, S.~Chakraborty, B.~Ray, and K.~Chang, ``{A Transformer-based
  Approach for Source Code Summarization},'' in \emph{Proceedings of the 58th
  Annual Meeting of the Association for Computational Linguistics, {ACL} 2020},
  pp. 4998--5007.

\bibitem{DBLP:journals/corr/abs-2111-08874}
J.~Cheng, I.~Fostiropoulos, and B.~W. Boehm, ``{GN-Transformer: Fusing Sequence
  and Graph Representation for Improved Code Summarization},'' \emph{CoRR},
  vol. abs/2111.08874, 2021.

\bibitem{DBLP:journals/corr/abs-2104-09340}
S.~Gao, C.~Gao, Y.~He, J.~Zeng, L.~Y. Nie, and X.~Xia, ``{Code Structure Guided
  Transformer for Source Code Summarization},'' \emph{CoRR}, vol.
  abs/2104.09340, 2021.

\bibitem{DBLP:conf/icsm/LeClairBM21}
A.~LeClair, A.~Bansal, and C.~McMillan, ``{Ensemble Models for Neural Source
  Code Summarization of Subroutines},'' in \emph{{IEEE} International
  Conference on Software Maintenance and Evolution, {ICSME} 2021}.\hskip 1em
  plus 0.5em minus 0.4em\relax {IEEE}, 2021, pp. 286--297.

\bibitem{DBLP:conf/emnlp/Shi0D0HZS21}
E.~Shi, Y.~Wang, L.~Du, H.~Zhang, S.~Han, D.~Zhang, and H.~Sun, ``{{CAST:}
  Enhancing Code Summarization with Hierarchical Splitting and Reconstruction
  of Abstract Syntax Trees},'' in \emph{Proceedings of the 2021 Conference on
  Empirical Methods in Natural Language Processing, {EMNLP} 2021}, pp.
  4053--4062.

\bibitem{Lopes}
``{Original Funcom Dataset},'' \url{http://www.ics.uci.edu/lopes/datasets/},
  2010.

\bibitem{shi2022we}
L.~Shi, F.~Mu, X.~Chen, S.~Wang, J.~Wang, Y.~Yang, G.~Li, X.~Xia, and Q.~Wang,
  ``{Are We Building on the Rock? On the Importance of Data Preprocessing for
  Code Summarization},'' \emph{arXiv preprint arXiv:2207.05579}, 2022.

\bibitem{DBLP:conf/acl/PapineniRWZ02}
K.~Papineni, S.~Roukos, T.~Ward, and W.~Zhu, ``{Bleu: A Method for Automatic
  Evaluation of Machine Translation},'' in \emph{Proceedings of the 40th Annual
  Meeting of the Association for Computational Linguistics}.\hskip 1em plus
  0.5em minus 0.4em\relax {ACL}, 2002, pp. 311--318.

\bibitem{lin2004rouge}
C.-Y. Lin, ``{ROUGE: A Package for Automatic Evaluation of Summaries},'' in
  \emph{Text summarization branches out}, 2004, pp. 74--81.

\bibitem{DBLP:conf/acl/BanerjeeL05}
S.~Banerjee and A.~Lavie, ``{{METEOR:} An Automatic Metric for {MT} Evaluation
  with Improved Correlation with Human Judgments},'' in \emph{Proceedings of
  the Workshop on Intrinsic and Extrinsic Evaluation Measures for Machine
  Translation and/or Summarization@ACL 2005}.\hskip 1em plus 0.5em minus
  0.4em\relax Association for Computational Linguistics, 2005, pp. 65--72.

\bibitem{DBLP:conf/icse/TangSLGHZ022}
Z.~Tang, X.~Shen, C.~Li, J.~Ge, L.~Huang, Z.~Zhu, and B.~Luo, ``{AST-Trans:
  Code Summarization with Efficient Tree-Structured Attention},'' in \emph{44th
  {IEEE/ACM} 44th International Conference on Software Engineering, {ICSE}
  2022}.\hskip 1em plus 0.5em minus 0.4em\relax {ACM}, 2022, pp. 150--162.

\bibitem{DBLP:journals/corr/KingmaB14}
D.~P. Kingma and J.~Ba, ``{Adam: {A} Method for Stochastic Optimization},'' in
  \emph{3rd International Conference on Learning Representations, {ICLR} 2015},
  2015.

\bibitem{DBLP:journals/corr/abs-1207-0580}
G.~E. Hinton, N.~Srivastava, A.~Krizhevsky, I.~Sutskever, and R.~Salakhutdinov,
  ``{Improving Neural Networks by Preventing Co-adaptation of Feature
  Detectors},'' \emph{CoRR}, vol. abs/1207.0580, 2012.

\bibitem{DBLP:conf/emnlp/WisemanR16}
S.~Wiseman and A.~M. Rush, ``{Sequence-to-Sequence Learning as Beam-Search
  Optimization},'' in \emph{Proceedings of the 2016 Conference on Empirical
  Methods in Natural Language Processing, {EMNLP} 2016}.\hskip 1em plus 0.5em
  minus 0.4em\relax The Association for Computational Linguistics, 2016, pp.
  1296--1306.

\bibitem{pytorch}
``{Pytorch Framework},'' \url{https://pytorch.org/}, 2016.

\bibitem{DBLP:conf/kbse/WanZYXY0Y18}
Y.~Wan, Z.~Zhao, M.~Yang, G.~Xu, H.~Ying, J.~Wu, and P.~S. Yu, ``{Improving
  Automatic Source Code Summarization via Deep Reinforcement Learning},'' in
  \emph{Proceedings of the 33rd {ACM/IEEE} International Conference on
  Automated Software Engineering, {ASE} 2018}, 2018, pp. 397--407.

\bibitem{DBLP:conf/kbse/SridharaHMPV10}
G.~Sridhara, E.~Hill, D.~Muppaneni, L.~L. Pollock, and K.~Vijay{-}Shanker,
  ``Towards {A}utomatically {G}enerating {S}ummary {C}omments for {J}ava
  {M}ethods,'' in \emph{{ASE} 2010, 25th {IEEE/ACM} International Conference on
  Automated Software Engineering}.\hskip 1em plus 0.5em minus 0.4em\relax
  {ACM}, 2010, pp. 43--52.

\bibitem{DBLP:conf/iwpc/MorenoASMPV13}
L.~Moreno, J.~Aponte, G.~Sridhara, A.~Marcus, L.~L. Pollock, and
  K.~Vijay{-}Shanker, ``{Automatic Generation of Natural Language Summaries for
  Java Classes},'' in \emph{{IEEE} 21st International Conference on Program
  Comprehension, {ICPC} 2013}.\hskip 1em plus 0.5em minus 0.4em\relax {IEEE}
  Computer Society, 2013, pp. 23--32.

\bibitem{DBLP:journals/tse/McBurneyM16}
P.~W. McBurney and C.~McMillan, ``{Automatic Source Code Summarization of
  Context for Java Methods},'' \emph{{IEEE} Trans. Software Eng.}, vol.~42,
  no.~2, pp. 103--119, 2016.

\bibitem{DBLP:conf/icse/HaiducAM10}
S.~Haiduc, J.~Aponte, and A.~Marcus, ``{Supporting Program Comprehension with
  Source Code Summarization},'' in \emph{Proceedings of the 32nd {ACM/IEEE}
  International Conference on Software Engineering - Volume 2, {ICSE}
  2010}.\hskip 1em plus 0.5em minus 0.4em\relax {ACM}, 2010, pp. 223--226.

\bibitem{DBLP:conf/wcre/HaiducAMM10}
S.~Haiduc, J.~Aponte, L.~Moreno, and A.~Marcus, ``{On the Use of Automated Text
  Summarization Techniques for Summarizing Source Code},'' in \emph{17th
  Working Conference on Reverse Engineering, {WCRE} 2010}.\hskip 1em plus 0.5em
  minus 0.4em\relax {IEEE} Computer Society, 2010, pp. 35--44.

\bibitem{DBLP:conf/iwpc/EddyRKC13}
B.~P. Eddy, J.~A. Robinson, N.~A. Kraft, and J.~C. Carver, ``{Evaluating Source
  Code Summarization Techniques: Replication and Expansion},'' in \emph{{IEEE}
  21st International Conference on Program Comprehension, {ICPC} 2013}.\hskip
  1em plus 0.5em minus 0.4em\relax {IEEE} Computer Society, 2013, pp. 13--22.

\bibitem{DBLP:conf/kbse/WongYT13}
E.~Wong, J.~Yang, and L.~Tan, ``{AutoComment: Mining Question and Answer Sites
  for Automatic Comment Generation},'' in \emph{2013 28th {IEEE/ACM}
  International Conference on Automated Software Engineering, {ASE}
  2013}.\hskip 1em plus 0.5em minus 0.4em\relax {IEEE}, 2013, pp. 562--567.

\bibitem{DBLP:conf/wcre/WongLT15}
E.~Wong, T.~Liu, and L.~Tan, ``{CloCom: Mining Existing Source Code for
  Automatic Comment Generation},'' in \emph{22nd {IEEE} International
  Conference on Software Analysis, Evolution, and Reengineering, {SANER}
  2015}.\hskip 1em plus 0.5em minus 0.4em\relax {IEEE} Computer Society, 2015,
  pp. 380--389.

\bibitem{DBLP:journals/jasis/DeerwesterDLFH90}
S.~C. Deerwester, S.~T. Dumais, T.~K. Landauer, G.~W. Furnas, and R.~A.
  Harshman, ``{Indexing by Latent Semantic Analysis},'' \emph{J. Am. Soc. Inf.
  Sci.}, vol.~41, no.~6, pp. 391--407, 1990.

\bibitem{DBLP:journals/cacm/SaltonWY75}
G.~Salton, A.~Wong, and C.~Yang, ``{A Vector Space Model for Automatic
  Indexing},'' \emph{Commun. {ACM}}, vol.~18, no.~11, pp. 613--620, 1975.

\bibitem{liu2018neural}
Z.~Liu, X.~Xia, A.~E. Hassan, D.~Lo, Z.~Xing, and X.~Wang,
  ``{Neural-machine-translation-based Commit Message Generation: How Far are
  We?}'' in \emph{Proceedings of the 33rd ACM/IEEE International Conference on
  Automated Software Engineering}, 2018, pp. 373--384.

\bibitem{DBLP:conf/acl/IyerKCZ16}
S.~Iyer, I.~Konstas, A.~Cheung, and L.~Zettlemoyer, ``{Summarizing Source Code
  using a Neural Attention Model},'' in \emph{Proceedings of the 54th Annual
  Meeting of the Association for Computational Linguistics, {ACL} 2016}.\hskip
  1em plus 0.5em minus 0.4em\relax The Association for Computer Linguistics,
  2016.

\bibitem{DBLP:conf/iwpc/HuLXLJ18}
X.~Hu, G.~Li, X.~Xia, D.~Lo, and Z.~Jin, ``{Deep Code Comment Generation},'' in
  \emph{Proceedings of the 26th Conference on Program Comprehension, {ICPC}
  2018}.\hskip 1em plus 0.5em minus 0.4em\relax {ACM}, 2018, pp. 200--210.

\bibitem{liu2021haconvgnn}
X.~Liu, D.~Wang, A.~Wang, Y.~Hou, and L.~Wu, ``{HAConvGNN: Hierarchical
  Attention Based Convolutional Graph Neural Network for Code Documentation
  Generation in Jupyter Notebooks},'' \emph{arXiv preprint arXiv:2104.01002},
  2021.

\bibitem{9031440}
W.~Wang, Y.~Zhang, Y.~Sui, Y.~Wan, Z.~Zhao, J.~Wu, P.~S. Yu, and G.~Xu,
  ``{Reinforcement-Learning-Guided Source Code Summarization Using Hierarchical
  Attention},'' \emph{IEEE Transactions on Software Engineering}, vol.~48,
  no.~1, pp. 102--119, 2022.

\bibitem{DBLP:journals/tosem/WangXLHWG21}
H.~Wang, X.~Xia, D.~Lo, Q.~He, X.~Wang, and J.~Grundy, ``{Context-aware
  Retrieval-based Deep Commit Message Generation},'' \emph{{ACM} Trans. Softw.
  Eng. Methodol.}, vol.~30, no.~4, pp. 56:1--56:30, 2021.

\bibitem{DBLP:conf/kbse/MuC0WW22}
\BIBentryALTinterwordspacing
F.~Mu, X.~Chen, L.~Shi, S.~Wang, and Q.~Wang, ``Automatic comment generation
  via multi-pass deliberation,'' in \emph{37th {IEEE/ACM} International
  Conference on Automated Software Engineering, {ASE} 2022, Rochester, MI, USA,
  October 10-14, 2022}.\hskip 1em plus 0.5em minus 0.4em\relax {ACM}, 2022, pp.
  14:1--14:12. [Online]. Available:
  \url{https://doi.org/10.1145/3551349.3556917}
\BIBentrySTDinterwordspacing

\bibitem{zhang2022survey}
H.~Zhang, H.~Song, S.~Li, M.~Zhou, and D.~Song, ``{A Survey of Controllable
  Text Generation Using Transformer-based Pre-trained Language Models},''
  \emph{arXiv preprint arXiv:2201.05337}, 2022.

\bibitem{pmlr-v70-hu17e}
Z.~Hu, Z.~Yang, X.~Liang, R.~Salakhutdinov, and E.~P. Xing, ``{Toward
  Controlled Generation of Text},'' in \emph{Proceedings of the 34th
  International Conference on Machine Learning}, ser. Proceedings of Machine
  Learning Research, vol.~70.\hskip 1em plus 0.5em minus 0.4em\relax PMLR,
  06--11 Aug 2017, pp. 1587--1596.

\bibitem{zhou2018emotional}
H.~Zhou, M.~Huang, T.~Zhang, X.~Zhu, and B.~Liu, ``{Emotional Chatting Machine:
  Emotional Conversation Generation with Internal and External Memory},'' in
  \emph{Proceedings of the AAAI Conference on Artificial Intelligence},
  vol.~32, no.~1, 2018.

\bibitem{xu2020megatron}
P.~Xu, M.~Patwary, M.~Shoeybi, R.~Puri, P.~Fung, A.~Anandkumar, and
  B.~Catanzaro, ``{MEGATRON-CNTRL: Controllable Story Generation with External
  Knowledge Using Large-scale Language Models},'' \emph{arXiv preprint
  arXiv:2010.00840}, 2020.

\bibitem{DBLP:journals/corr/abs-1909-05858}
N.~S. Keskar, B.~McCann, L.~R. Varshney, C.~Xiong, and R.~Socher, ``{{CTRL:}
  {A} Conditional Transformer Language Model for Controllable Generation},''
  \emph{CoRR}, vol. abs/1909.05858, 2019.

\end{thebibliography}

\end{document}